\documentclass[num-refs]{wiley-article}

\usepackage{siunitx}

\usepackage{amssymb}
\usepackage{xcolor,subfigure}
\usepackage{amsmath,adjustbox,parskip,natbib}
\usepackage{hyperref}
\usepackage{booktabs}
\usepackage{rotating}

\papertype{Original Article}

\title{Time-Series Foundation AI Model for Value-at-Risk Forecasting}

\author[1]{Anubha Goel}
\author[2]{Puneet Pasricha}
\author[1]{Juho Kanniainen}

\affil[1]{Department of Computing Sciences, Financial Computing and Data Analytics Group, Tampere University, Tampere, 33100, Finland}
\affil[2]{Department of Mathematics, IIT Ropar, Punjab, 140001, India}

\corraddress{Juho Kanniainen, Department of Computing Sciences, Financial Computing and Data Analytics Group, Tampere University, Tampere, 33100, Finland}
\corremail{juho.kanniainen@tuni.fi}

\fundinginfo{Anubha Goel, Department of Computing Sciences, Grant/Award Number: 101150609 (Marie Skłodowska-Curie Action), Project: HiddenTipChains }

\runningauthor{Anubha Goel et al.}

\begin{document}

\begin{frontmatter}
\maketitle

\begin{abstract}
This study is the first to analyze the performance of a time-series foundation AI model for Value-at-Risk (VaR), which essentially forecasts the left-tail quantiles of returns. Foundation models, pre-trained on diverse datasets, can be applied in a zero-shot setting with minimal data or further improved through finetuning. We compare Google’s TimesFM model to conventional parametric and non-parametric models, including GARCH and Generalized Autoregressive Score (GAS), using 19 years of daily returns from the S\&P 100 index and its constituents. Backtesting with over 8.5 years of out-of-sample data shows that the fine-tuned foundation model consistently outperforms traditional methods in actual-over-expected ratios. For the quantile score loss function, it performs comparably to the best econometric model, GAS. Overall, the foundation model ranks as the best or among the top performers across the 0.01, 0.025, 0.05, and 0.1 quantile forecasting. Fine-tuning significantly improves accuracy, showing that zero-shot use is not optimal for VaR.

\end{abstract}
\end{frontmatter}
\section{Introduction}
\label{sec:introduction}

Although machine learning (ML) has significantly advanced financial market research---such as asset pricing, volatility forecasting,  limit order book dynamics, and optimal hedging \citep{gu2020empirical,christensen2023machine,tran2018temporal,mikkila2023empirical}, the risk management literature still largely relies on traditional econometric techniques in time series forecasting, especially in the area of Value-at-Risk (VaR). VaR is essentially about forecasting the left-tail quantiles (e.g., the 5th or 1st percentile) of return distributions to quantify potential losses over a given time period. VaR forecasts are often based on daily return observations \citep[for example, see][based solely on historical daily returns]{wang2022forecasting,patton2019dynamic,taylor2019forecasting,berger2021assessing,wang2024bayesian,storti2023modeling}, and even 20 years of data would yield only around 5,000 observations. This is quite limited, especially considering that DL models often contain millions of parameters that need to be learned. 

However, a solution to this problem is currently underway. Recently, ML has experienced a paradigm shift with the emergence of foundation models, which are large-scale, general-purpose pretrained, and often fine-tunable models on diverse datasets spanning various data distributions. 
Although the foundation models are based on deep learning and transfer learning principles, their scale leads to the emergence of new capabilities \citep{bommasani2021opportunities}. 
In 2023--2024, following major advancements in large language models (LLMs), several pre-trained time-series foundation models have emerged, including Google's TimesFM \citep{das2024decoder}, Time-GPT \citep{garza2024timegpt}, and LagLlama \citep{rasul2023lag}. 

AI-driven advancements in financial risk modeling have opened doors to new methodologies for assessing complex financial data \citep{tran2019data,guijarro2021deep,chen2024deep}. Historically, methods such as GARCH-based models have been effective in capturing dependencies and clustering of volatility in market data \citep{hoga2023monitoring}. However, recent approaches, such as Generative Adversarial Networks (GANs), Variational Autoencoders (VAEs), and Conditional GANs (CGANs), enhance distribution forecasting by generating realistic synthetic data \citep{nareklishvili2023generative, athey2021using, hofert2022multivariate}. Such models provide flexible, non-parametric approaches for data structures that traditional econometric tools may struggle to approximate. \cite{ericson2024deep} presents a comparative review showing that deep-generative methods outperform traditional benchmarks in modeling complex financial dynamics for VaR forecasting. These advancements highlight the potential of AI-driven approaches in risk forecasting. This study stands out from previous research by adopting a time-series foundation model, pre-trained on diverse datasets, to forecast VaR. Unlike traditional supervised ML models trained for specific tasks, time-series foundation models are pre-trained on vast datasets spanning billions of time points. A single pre-trained model can be applied across a wide range of domains, from traffic forecasting to stock price predictions, offering quick and cost-effective solutions. The term `foundational' emphasizes their flexibility, as they can be adapted for different tasks or domains with minimal additional training. The foundation models can be used with or without fine-tuning.

We implement a pre-trained time-series foundation model, specifically Google's TimesFM \citep{das2024decoder}, for VaR forecasting and compare it with existing state-of-the-art parametric and semi-parametric approaches, including GARCH model, Generalized Autoregressive Score (GAS) one-factor model, and empirical quantile estimates \citep{wang2022forecasting}. The key question is whether the pre-trained foundation model yields results comparable to, or even capable of outperforming, existing econometric methods. An additional objective is to assess the extent to which fine-tuning improves the accuracy of VaR predictions.

Rather than identifying the best foundation model, which is subject to frequent updates and new developments, this study aims to address the fundamental question of whether state-of-the-art parametric and semi-parametric methods can be outperformed. While several foundation models have been recently introduced, this study focuses on TimesFM due to its robust performance, flexibility, and openness, which align well with the methodological and practical needs of our research. Other contributing factors in the selection include its open-source availability\footnote{\url{https://github.com/google-research/TimesFM}.}, modifiability, and free accessibility, ensuring the results are fully replicable.\footnote{The codes of this paper can be found at \url{https://github.com/Anubha0812/TimesFM-for-Value-at-Risk}.} To be more precise, unlike proprietary models such as TimeGPT, TimesFM is open and fine-tuneable, allowing researchers to adapt the model to domain-specific datasets. TimesFM is pre-trained on a large and diverse collection of public and synthetic time-series datasets, enabling it to deliver strong zero-shot performance across a variety of forecasting tasks. One of TimesFM's key advantages is its flexibility at inference time: it supports varying input lengths, forecast horizons, and time granularities, making it highly adaptable across different forecasting scenarios. TimesFM provides probabilistic forecasts, delivering quantile estimates (deciles) along with the mean forecast. These quantile heads can be fine-tuned as well making it more suitable for the current study. Furthermore, the authors of TimesFM have empirically demonstrated that their model outperforms TimeGPT in forecasting accuracy, adding further motivation for its selection. Finally, TimesFM has been integrated into GIFT-Eval \footnote{\url{https://huggingface.co/spaces/Salesforce/GIFT-Eval}}, a comprehensive benchmarking framework for time-series models. Within this framework, it ranks at the top based on aggregated Mean Absolute Scaled Error (MASE) and Continuous Ranked Probability Score (CRPS), demonstrating its superior generalization capabilities.

We conduct an extensive empirical evaluation of the performance of the foundation time-series model using daily return data on the 91 constituents of the S\&P 100 index over 19 years from January 2005 September 2023. The first ten years of data is used for estimating (learning) model parameters, and the remaining 9 years for out-of-sample testing the performance of various models. For each constituent, we calculate 1-day VaR on a daily basis. The foundation model is applied both with and without fine-tuning to understand how much the results improve with fine-tuning, which has certain computational cost. The fine-tuning is performed on a daily basis using a rolling window technique.


Our main finding is that, in terms of the actual-over-expected ratio, the fine-tuned time-series foundation model outperforms existing econometric approaches for VaR forecasting. Specifically, the fine-tuned foundation model consistently demonstrates superior performance, on average, across all 91 constituents of the S\&P 100 included in the analysis. Regarding the quantile score loss function, which is the second backtesting metric, the fine-tuned foundation model achieves performance comparable to the best econometric approach, the GAS model. Overall, the foundation model is either the top performer or among the best in forecasting VaR across the 0.01, 0.025, 0.05, and 0.1 VaR levels. Moreover, we found that fine-tuning is practically necessary for the foundation model; using it as pre-trained in zero-shot settings results in significantly worse performance compared to the best econometric approaches. We consider this as a remarkable result on the performance of Google's TimesFM model as this model is not explicitly designed for this purpose and was not even pre-trained with stock return data. This problem is statistically challenging due to its focus on heteroscedastic processes with time-varying conditional volatility. In this light, the paper is valuable not only for financial risk management, but also as a benchmark for testing AI time-series models against state-of-the-art statistical approaches on a highly complex issue.


The paper is structured as follows: In Section 2, we describe the time-series foundation model, Google's TimesFM, and the benchmark models against which it is compared. Section 3 details the data and experimental setup for the models and fine-tuning specifications for VaR forecasting. The evaluation metrics and backtesting procedures are outlined in Section 4. Section 5 presents an analysis of the empirical results, examining the performance of the foundation model relative to traditional econometric approaches. Finally, the findings have several implications beyond Value-at-Risk forecasting techniques, which, along with the limitations of our analysis, will be discussed at the end of the paper in Section \ref{sec:conclusion_discussion}.

\section{Models}\label{models}

\subsection{Time-Series Foundation Models}

The time-series foundation models share similarities with large language models (LLMs) as they rely on transformer-based architectures to capture long-range dependencies and patterns within sequential data. While LLMs are trained to predict the next or missing part of a text sequence, foundational time-series models process continuous time points as individual tokens. This approach allows them to handle time-series data step by step. 

TimesFM model \citep{das2024decoder} is a decoder-only transformer architecture, similar to GPT models. In contrast to encoder-decoder models, where the encoder processes the input sequence and the decoder generates output based on the encoded input, decoder-only models focus solely on output generation, processing the input through successive decoder layers. This allows for faster training and inference times compared to encoder-decoder models. Decoder-only models like are highly advantageous for zero-shot forecasting, where minimal fine-tuning is required. On the other hand, the advantage of the encoder-decoder architecture, which is used, for example, in TimeGPT \citep{garza2024timegpt}, is that it can handle more complex inputs and exogenous variables. However, in this paper, we use univariate and well-structured time-series data on returns, and VaR is calculated separately for each security, and for that reason, the encoder component does not add significant value. 

\begin{figure}[ht!]
    \centering
    \includegraphics[width=0.7\linewidth,angle=-90]{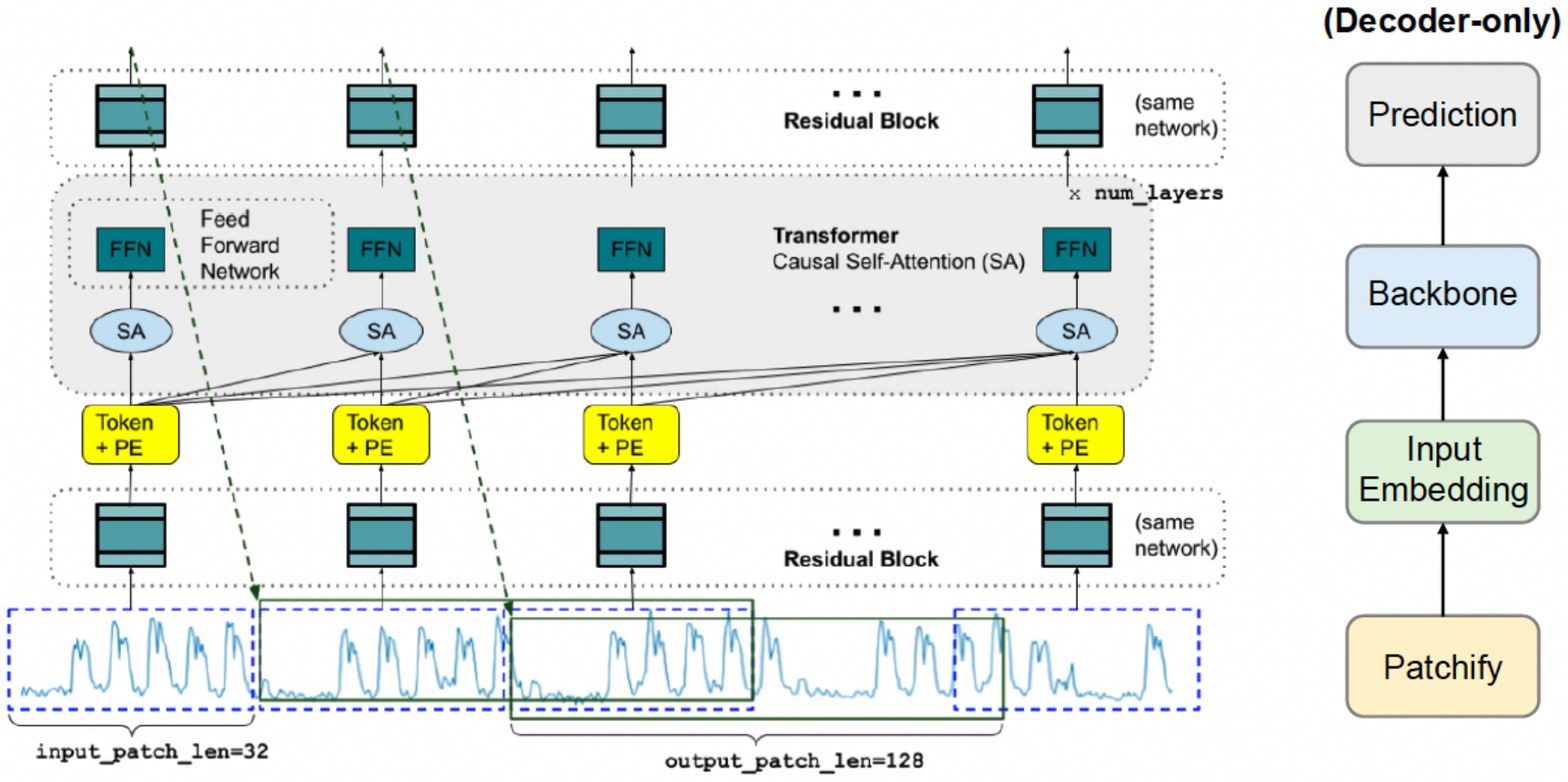}
    \caption{TimesFM structure Note. From \cite{das2024decoder}. A Decoder-Only Foundation Model for Time-Series Forecasting. Proceedings of the Forty-First International Conference on Machine Learning (ICML). Copyright 2024 by the Authors. The figure presents the architecture of TimesFM, a time-series foundation model built on a decoder-only transformer framework. }
    \label{fig:tfm}
\end{figure}

The architecture of TimesFM is optimized for long-horizon forecasting through a decoder-only transformer framework that employs a strategic patching mechanism. Time-series data is segmented into non-overlapping patches, which serve as analogues to tokens in language models. This patch-based representation not only improves computational efficiency by reducing input sequence length, but also enhances the model’s capacity to capture temporal dependencies over extended horizons. Each input patch is transformed via a residual MLP block and enriched with positional encodings before being processed by stacked multi-head causal self-attention layers. Unlike traditional autoregressive models that forecast one step at a time, TimesFM predicts longer output patches in parallel, significantly reducing the number of autoregressive iterations needed during inference. This architectural design makes the model highly adaptable to varying context lengths, forecasting horizons, and temporal granularities. Furthermore, the use of dynamic patch masking during training ensures that the model generalizes well across different time series patterns and lengths, making it a robust choice for zero-shot forecasting tasks in financial applications (see Figure \ref{fig:tfm} for a schematic overview of the TimesFM architecture).

In pretraining, TimesFM was trained on a vast corpus comprising both real-world time-series data such as Google trends, Wiki Pageviews, Synthetic Data, and Electricity, Traffic, and Weather datasets used in M4 competition \citep{makridakis2020m4,spiliotis2020forecasting}. However, financial data was not used, and this aspect makes the application of the present paper particularly interesting.
This corpus, although smaller in scale compared to text-based models, spans around 100 billion time points, providing the diversity needed for the model to generalize well across domains and forecasting tasks. 
Compared to the state-of-the-art LLMs, TimesFM time-series foundation model is much smaller, with 200M parameters and pretraining data consisting of around 100B timepoints. Despite its smaller scale, it was reported to achieve relatively high zero-shot forecasting accuracy.

\subsection{Benchmark Models for VaR}

To evaluate the performance of the foundation time series models for estimating Value-at-Risk, we benchmark it against several established models. These include both traditional and advanced methods, each with its own set of assumptions and computational characteristics. The benchmark models are as follows:

\begin{itemize}
\item Rolling Window Quantile Estimation:

One of the most widely used approaches for VaR forecasting is the rolling window method, which computes the risk measure based on historical data over a fixed time window:
\begin{equation*}
    \widehat{\text{VaR}}_t = \text{Quantile} \left\{ r_i \right\}_{i=t-m}^{t-1},
\end{equation*}
where $\text{Quantile} \left\{ r_i \right\}_{i=t-m}^{t-1}$ represents the sample quantile of the variable $r_i$ over the period $i \in [t-m, t-1]$. The parameter $m$ denotes the size of the rolling window. While this approach is straightforward and easy to implement, it does not account for the dynamic nature of financial markets, leading to potential inaccuracies during periods of high volatility.

\item ARMA-GARCH Models \citep{angelidis2004use}:

To better capture the time-varying properties of financial returns, we employ ARMA-GARCH models. These models estimate the conditional mean and variance of returns, incorporating various distributional assumptions for the standardized residuals. The model is specified as
\begin{equation*}
    r_t = \mu_t + \sigma_t \eta_t, \text{\quad}    \eta_t \sim \text{iid } \mathbb{G}_\eta (0, 1).
\end{equation*}
In this framework, $\mu_t$ represents the conditional mean, typically modeled using an ARMA process, while $\sigma_t^2$ denotes the conditional variance, modeled using a GARCH process. The standardized residuals $\eta_t$ follow a distribution $\mathbb{G}_\eta$ with mean zero and variance one.
The VaR for these models is calculated as 
\begin{equation*}
    v_t = \mu_t + \sigma_t a_\alpha, \quad \text{where } a_\alpha = \mathbb{G}_\eta^{-1} (\alpha).
\end{equation*}
Here, $a_\alpha$ is the $\alpha$-quantile of the distribution $\mathbb{G}_\eta$.
Different assumptions for $\mathbb{G}_\eta$ yield various model specifications, including:
\begin{enumerate}
    \item Normal GARCH: This specification assumes that the residuals follow a normal distribution:
\begin{equation*}
    \eta_t \sim \text{iid } N (0, 1).
\end{equation*}
This is the most basic assumption and is often insufficient for capturing the heavy tails observed in financial return distributions.
\item Student's t-GARCH: To account for the heavy tails and potential skewness in the return distribution, a skewed t-distribution is used for the residuals:
\begin{equation*}
    \eta_t \sim \text{iid Skew } t (0, 1, \nu, \lambda).
\end{equation*}
This model provides greater flexibility in capturing the tail behavior of financial returns, which is crucial for accurate risk estimation.

\item Empirical Distribution Function: As a nonparametric alternative, we use the Empirical Distribution Function (EDF) method \citep{manganelli2004comparison}. This approach estimates the distribution of standardized residuals $\eta_t$ from historical data, avoiding any parametric assumptions. 
\end{enumerate}

\item One factor Gas model \citep{patton2019dynamic}: This model belongs to the class of CAViaR models \citep{taylor2005generating}, where the evolution of the risk measure is directly parameterized without the need to specify a full conditional distribution for asset returns.\footnote{GAS model was originally proposed in \cite{creal2013generalized}. \cite{patton2019dynamic} proposed a dynamic GAS model employing the class of loss functions proposed in \cite{fissler2016}, called FZ loss functions. In this paper, we adopt the GAS model from Section 2.3 in \cite{patton2019dynamic}.} The evolution of VaR for the Gas 1 factor model is given 
\begin{equation*}
    v_t = a \exp(\kappa_t),
\end{equation*}
where $a < 0$ is a scaling parameter and $\kappa_t$ represents the dynamic risk factor driving the VaR process. The evolution of $\kappa_t$ is given by the updating mechanism
\begin{equation*}\begin{split}
    \kappa_t &= \omega + \beta \kappa_{t-1} \\
    &+ \gamma \frac{1}{b\exp(\kappa_{t-1})} \left( \frac{1}{\alpha} \mathbf{1}\{r_{t-1} \leq a \exp(\kappa_{t-1}) \} r_{t-1} - b \exp(\kappa_{t-1}) \right),
\end{split}\end{equation*}
where $b < a < 0 < \gamma $ and we set $\omega=0$ as in \cite{patton2019dynamic}

\end{itemize}
\section{Data and Experiments}

We now apply the models discussed in the last section to the forecasting of VaR for daily returns on the S\&P 100 index along with its 91 constituents. 
The data used in our analysis was obtained from Thomson Reuters Datastream, covering a period of approximately 19 years, from January 2005 to September 30, 2023, resulting in 4,876 observations. Throughout this period, data on every day was available for 91 stocks.\footnote{The S\&P100 Global Index typically comprises 100 companies; however, as of October 1, 2023, it included 102 stocks (as Roche Holding AG and GOOGLE have two classes of shares). However, due to the unavailability of data for certain stocks, specifically Synlogic Inc (SYBX UR Equity), Alphabet Inc (GOOG UW Equity), Seven \& i Holdings Co Ltd (3382 JT Equity), Texas Instruments Inc (TXN UW Equity), Shell PLC (SHEL LN Equity), Philip Morris International Inc (PM UN Equity), Mastercard Inc (MA UN Equity), Broadcom Inc (AVGO UW Equity), Engie SA (ENGI FP Equity), PepsiCo Inc (PEP UW Equity), and Honeywell International Inc (HON UW Equity)—the analysis was conducted on the remaining 91 constituents}. The daily return for $i$-th stock at $t$-th day is calculated as $r_{it}=\frac{p_{it}-p_{it-1}}{p_{it-1}};$ $i=1,\ldots, n, \ t=1,\ldots, T,$  where $p_{it}$ and $p_{it-1}$ are respectively, the closing prices at $t$-th and $(t-1)$-th day. We divide the data into two consecutive, non-overlapping samples while maintaining the temporal ordering of the data. We use the first ten years of data, i.e., January 2005 to December 2014, for estimating model parameters and the remaining 9 years, i.e., January 2015 to September 2023, for out-of-sample testing the performance of various models.

For a comprehensive view of the data, Table \ref{tab:summary_stats} reports the summary statistics for key risk and return metrics of the 91 constituent stocks and the index. Specifically, it reports the mean, median, standard deviation, minimum, and maximum values for the Mean, Standard Deviation, Skewness, Kurtosis, VaR, and CVaR of these assets. From skewness and kurtosis values, we observe that the distribution of these returns are fat tailed and are skewed to some extent, thus, agreeing with the stylized facts of empirical returns. Also, the mean VaR across 91 stocks is less than that of the index, showing that index is less riskier compared to the constituent stocks.

\begin{table}[ht!]
  \centering
  \begin{tabular}{l|ccccc|c}
          & \multicolumn{5}{c|}{Constituents}       & \multicolumn{1}{c}{Index} \\
          \hline
          & \multicolumn{1}{c}{Min} & \multicolumn{1}{c}{Max} & \multicolumn{1}{c}{Median} & \multicolumn{1}{c}{Mean} & \multicolumn{1}{c|}{Std-Dev} &  \\
          \hline
    Mean  & -0.013 & 0.157 & 0.031 & 0.038 & 0.029 & 0.024 \\
    Std-Dev & 0.106 & 0.311 & 0.178 & 0.184 & 0.048 & 0.107 \\
    Skewness & -0.502 & 5.187 & 0.193 & 0.302 & 0.700 & -0.252 \\
    Kurtosis & 3.139 & 155.367 & 9.513 & 13.093 & 18.226 & 10.292 \\
    VaR (99\%) & -8.429 & -2.867 & -4.848 & -4.991 & 1.286 & -3.249 \\
    VaR (97.5\%) & -5.907 & -2.128 & -3.519 & -3.592 & 0.873 & -2.195 \\
    VaR (95\%) & -4.433 & -1.521 & -2.641 & -2.674 & 0.635 & -1.578 \\
    VaR (90\%) & -3.093 & -1.048 & -1.844 & -1.848 & 0.435 & -1.069 \\
  \end{tabular}%
\caption{Summary statistics for Mean, Standard Deviation, Skewness, Kurtosis, VaR, and CVaR of the percentage returns of S\&P 100 index and its 91 constituents for data from January 2005 to
September, 2023.}
\label{tab:summary_stats}
\end{table}%

\subsection{Experiment Setup}

We implement the recently introduced pre-trained time-series model, Google's TimesFM, for forecasting Value-at-risk and compare its performance against state-of-the-art parametric and non-parametric approaches, namely, GARCH model with the standardized residual modeled as empirical distribution function (G-EDF), normal distribution (G-N) and Student's-t (G-t) distribution, one factor Generalized AutoRegressive Score model (GAS) and rolling window method (Historical). For each of the models, G-EDF, G-N, G-t, one factor GAS, and Historical, we obtained the VaR forecasts at levels 1\%, 2.5\%, 5\%, and 10\%.

For GARCH and GAS models, we followed \cite{patton2019dynamic} and used January 2005 to December 2014 as a training period for estimating model parameters, which were then used to forecast over the out-of-sample test period from the 2nd January 2015 to the 19th September 2023, thus resulting in a sequence of 2,268 VaR forecasts. For the rolling window technique  with the Historical Method, we used the data of 512 days preceding the first test day (i.e., 2nd January 2015), and forecasted the VaR at different levels. We then shift the training sample by one day forward in time to include more recent data and exclude the oldest data point such that a fixed size of the rolling window is maintained. At each rolling step, we obtain the prediction for the next day using the most recent 512 observations, thus resulting in a sequence of 2,268 VaR forecasts for each quantile level.

TimesFM is originally trained on a large number of public and synthetic datasets and exhibits robust out-of-the-box zero-shot performance in comparison to the accuracy of various state-of-the-art forecasting models specific to individual datasets under consideration. The model is flexible enough to forecast with different input length, prediction length and time granularities at inference time. The pre-trained model returns the point forecasts along with deciles, i.e., the quantiles\footnote{As per Google'e TimesFM GitHub page, ``We experimentally offer quantile heads but they have not been calibrated after pretraining."} at $10\%, 20\%,\ldots,90\%$ levels. We experimented with different prediction horizons with a fixed input length of 2 years (512 days). Specifically, we set the prediction horizon to 1 day, 21 days and 63 days and denote these models as PT1, PT21, and PT63. For PT1, the input window shifts daily with the prediction points, while for PT21 and PT63, we use a fixed set of input observations for daily forecasts over the subsequent 21 and 63 days, respectively. For example, with PT21 (or PT63), we estimated one-day VaRs over a 21-day (or 63-day) period starting from January 2, 2015, using the most recent 512 observations. After this, we shifted the training data forward by 21 days (or 63 days) to incorporate more recent data and applied the pre-trained model for the next 21 (or 63) days, generating one-day VaRs forecasts over this period. The variation in prediction horizon allows us to examine the model's ability to generalize across forecasting horizons of varying lengths, i.e., short and long horizons. However, as mentioned above, the pre-trained models, PT1, PT21, and PT63, only produced VaR forecast at 90\% level.

To better align the model with our objective, we conducted a fine-tuning procedure using the stock returns of S\&P100 and its 91 constituents. We adopted the code ``finetuning" in TimesFM library. More specifically, fine-tuning was carried out using a PatchedDecoderFinetuneModel with specific adjustments tailored for our task. During fine-tuning, we applied linear probing where only the core layer parameters were updated but the transformer layers were kept fixed. To prevent gradient explosion, we applied gradient clipping with a threshold of 100. The optimization process employed the Adam optimizer, that followed a cosine schedule for learning rate that started at $1\times 10^{-3}$ and decayed to $1\times 10^{-4}$ over 40,000 steps. Further, an exponential moving average decay of 0.9999 was also included in the optimization. The training loop ran for up to 100 epochs, with a provision of early stopping tied to the loss on the validation set and also a patience level was set to 5 epochs without improvement. Throughout the fine-tuning procedure, check-points were saved and finally the best model was restored for evaluation purposes. 

In the fine-tuning process, we divided the first ten-year period, from January 2005 to December 2014, into two subsets: the training set from January 2005 to December 2011, and the validation set from January 2012 to December 2014. The validation data was used for hyperparameter optimization 
and to track the model's performance after every epoch. That is, validation data is used to tune hyper-parameters by evaluating the performance on validation set during training, and selecting the best model check-point, that is, the one with the lowest validation loss. With a fixed input length of 512, we fine-tuned the pre-trained model separately for three different prediction horizons. Specifically, we obtained the fine-tuned model FT1 using a prediction horizon of 1 day, FT21 with a horizon of 21 days, and FT63 with a horizon of 63 days. Overall, three fine-tuned models were generated, denoted as FT1, FT21, and FT63, corresponding to the different prediction horizons. With fine-tuning, we re-trained the model for the quantile levels 0.01, 0.025, 0.05, and 0.1 as the pre-trained models used quantiles 0.1, 0.2,... 0.9 only.

\section{Evaluation metrics}

In financial risk management, VaR is a widely used measure that estimates the potential loss in the value of a stock over a given time period, with a certain confidence level. However, VaR forecasts are erroneous and hence often biased due to various sources of errors such as model errors, sampling errors, etc. Therefore, to ensure that VaR models are reliable, it is crucial to perform rigorous backtesting. Backtesting is the process of comparing the VaR predictions against actual observed outcomes to determine the accuracy of the model. Since a VaR value is fundamentally unobservable, we need to judge the performance of a model by checking whether these estimates are consistent with the realized returns or not, given the confidence level at which these estimates were obtained. Thus, VaR backtesting is critical for evaluating the accuracy of the VaR forecasts.

Given the daily out-of-sample VaR forecasts $\{q_t^{(m)}\}_{t=1}^T$ by model $m$ over the period $t=1$ to $t=T$, the simplest approach to access the accuracy of the VaR forecasts is to compare the number of violations against the expected number of violations. Here, a violation is an instance when the actual return is below the VaR forecast. Therefore, we calculate the actual over expected (AE) ratio, defined as 
\begin{equation}
   \text{AE} = \frac{\text{Actual Violations (A)}}{\text{Expected Violations (E)}}
\end{equation}
where the expected violations are calculated as \(E =T \times \alpha\), with \(T\) being the total number of observations and \(\alpha\) the confidence level of the VaR model. Clearly, an AE ratio of 1 is desirable, as it indicates that the model is accurately calibrated. On the other hand, a value $>1$ ($<1$) suggests the underestimation (overestimation) of the risk, both of which are problematic in the sense that an underestimate leads to unexpected negative returns, while the latter leads to high capital requirements, thus, both implying capital losses. Alternatively, a model is deemed good if $|1 - \text{AE}|$ is close to zero.

An alternative to AE ratio is the proportion of failure test (POF), also known as the unconditional coverage test (UC). which statistically evaluates whether the actual number of VaR violations matches the expected number of violations over time, based on the chosen confidence level. For instance, it calculates the failure rate defined as the proportion of the number of violations ($A$) to the total number of observations $T$ and examines its consistency with the model's confidence level $\alpha$. More specifically, let $I_t$ denotes the indicator function that the realized return was less than the VaR estimate, i.e.,
\begin{equation*}
I_t=\begin{cases}
1 &\ r_t < q_t\\
0 ,&\ \text{otherwise},
\end{cases}    
\end{equation*}
where $r_t$ is the realized return at time $t$. Thus, $A=\sum_{t=1}^TI_t$ denotes the number of violations in the sample. From \cite{kupiec1995techniques}, $A$ follows a binomial distribution, i.e., $A\sim B(T,\alpha)$, thus, under the null that $A/T=\alpha$, the relevant likelihood ratio statistic is
\begin{equation*}
2\ln\bigg[\bigg(1-\frac{A}{T}\bigg)^{T-A}\bigg(\frac{A}{T}\bigg)A\bigg]-2\ln[(1-\alpha)^{T-A}\alpha^A],  
\end{equation*}
which, asymptotically, follows a chi-squared distribution with one degrees of freedom.

As stated in \cite{kupiec1995techniques}, the UC test is capable of rejecting a model for both low and high failures, but has poor power in general. Further, the UC test tests the number of violations, while it is crucial to test whether these losses are spread evenly over time or not. The Conditional Coverage (CC) test, proposed by \cite{christoffersen1998evaluating}, does the same, that is, it tests whether the violations are randomly distributed over time and not clustered. That is, CC test takes into account any conditionality in VaR forecasts: the model must respond to clustering events such as if volatilities are high in some periods and low in others. The test statistics is given by
\begin{equation*}
-2\ln[(1-\alpha)^{T-A}\alpha^A]+2\ln[(1-\pi_{01})^{n_{00}}\pi_{01}^{n_{01}}(1-\pi_{11})^{n_{10}}\pi_{11}^{n_{11}}]    
\end{equation*}
that follows a chi-squared distribution with 2 degrees of freedom and
\begin{equation*}
\pi_{ij}=\frac{n_{ij}}{\sum_{j}n_{ij}}   
\end{equation*}
with $n_{ij}$ denoting the number of observations with value $i$ followed by $j$, and $\pi_{ij}$ denoting the corresponding probabilities. Finally, the Dynamic Quantile (DQ) test provides a comprehensive evaluation of the VaR model. It goes beyond just unconditional coverage and independence, and checks for any dynamic misspecification in the model, for instance, checks for any dependencies or patterns in forecasted errors that model failed to capture. These tests serve distinct but complementary purposes in validating the robustness of the VaR model.

In the evaluation of VaR forecasts, the quantile score is a commonly used metric and is now standard in the VaR literature \citep{ehm2016quantiles, gneiting2011making, taylor2019forecasting,bams2017volatility}. The above mentioned statistical tests are good to evaluate the statistical adequacy of VaR forecasts, that is, a model for VaR forecasting is deemed to be an adequate model if we fail to reject the null hypothesis. However, these tests cannot conclude whether an `adequate' model outperforms another `adequate' model. Scoring functions allow us to do an honest assessment of VaR as a perfect VaR estimate would always minimize the scoring function. The quantile score  $QS_t$ can be defined as follows:
\begin{equation}
   QS_t = \left(\alpha - \mathbf{1}\{ r_t < q_t \}\right) \times (r_t - q_t).
\end{equation}

For every instant $t$, a model's score increases either by (a) $\alpha-1$ fraction of the distance between the realized return and the VaR forecast, if realized return is smaller than the VaR forecast for the same period, or (b) by $\alpha$ fraction of the distance otherwise. This scoring rule is consistent for quantile estimation, meaning that it achieves its minimum value when the quantile forecast $q_t$ correctly estimates the true quantile of the distribution. Therefore, models that produce lower average quantile scores in the out-of-sample period are considered superior. Finally, to compare the performance of two models, we use the Diebold–Mariano (DM) test \citep{diebold2002comparing} with 
\begin{equation*}
H_0 : \text{quantile score}_1 = \text{quantile score}_2, 
\end{equation*}
against a one-sided alternative 
\begin{equation*}
H_1 : \text{quantile score}_1 > \text{quantile score}_2, 
\end{equation*}
where $\text{quantile score}_i$ denotes the quantile score from model $i$, i=1,2.

\section{Results}
\subsection{Actual over Expected Ratio}

Table \ref{table_ae} reports the summary statistics over the out-of-sample period from January 2015 to September 2023, across 92 stocks, of the $|1 - \text{AE}|$ values for the VaR forecasts at different levels, and the corresponding models in the out-sample period. Specifically, we report the minimum, mean, median, maximum, and standard deviation across all the stocks. Additionally, we report the count of stocks for which each of the considered models was the best (achieved lowest value of $|1-AE|$) or was within top two models (1st-2nd best). Note that, as mentioned before, the VaR forecasts for TimesFM pretrained model are only available at $10\%$, there we have three extra models in the last block of Table \ref{table_ae}.

Several interesting findings emerge. First, the fine-tuned TimesFM is superior to the pre-trained TimesFM model. For instance, models FT1, FT21 and FT63 attained the lowest $|1-AE|$ value in 22, 10 and 1 assets respectively, in comparison to their pretrained counterparts that attained lowest value in 1, 9 and 0 assets. Second, for the fine-tuned TimesFM model, the mean value of $|1-AE|$ is better when the prediction length is large for VaR at $1\%$ and $2.5\%$, whereas it is better for smaller prediction length when forecasting VaR at $5\%$ and $10\%$ levels. Third, for each VaR level, fine-tuned TimesFM model FT1 performed fairly well in comparison to the benchmark models. We also performed a two-sample $t$-test to check whether the mean, over the 92 stocks, of $|1-AE|$ from one model is significantly different from that from the other model. More specifically, we performed two-sample $t$-test for each pair of model where we test 
\begin{equation*}
H_0 : |1-AE|_i = |1-AE|_j, 
\end{equation*}
against a one-sided alternative 
\begin{equation*}
H_1 : |1-AE|_i > |1-AE|_j, 
\end{equation*}
where \( i \) denotes the model in the selected row, whereas \( j \) denotes the model in the selected column. Further, the formatting in the Tables are as follows: *, ** and *** denote whether the $t$-test of equal predictive accuracy is rejected at the 5\%, 2.5\% and 1\% level of significance. Table \ref{table_aetest} reports the out of sample results to $t$-tests. We can observe that the fine-tuned models, in particular, FT1 and FT21, perform remarkably well across all quantile levels, depicted by failure to reject the null hypothesis in the first and second rows of each block. Further, we observe that FT1 outperforms GAS model in forecasting VaR(1\%), VaR(5\%) and VaR(10\%) at 97.5\%, 95\% and 99\% level of significance respectively, and performs atleast as good as GAS model for forecasting Var(2.5\%) (refer rows corresponding to GAS model). In addition, we observe that selecting a short horizon as prediction length performs better in predicting VaR. What's more, the performance of pre-trained models is not so exciting compared to their fine-tuned counterparts. These findings show the potential of time-series foundation models in forecasting value-at-risk.

\begin{sidewaystable}[htbp]
  \centering
\resizebox{.90\textwidth}{!}{
  \begin{tabular}{l|ccccccc|ccccccc}
          & \multicolumn{7}{c|}{VaR (1\%)} & \multicolumn{7}{c}{VaR (2.5\%)}\\
          \hline
          & \multicolumn{1}{c}{Min} & \multicolumn{1}{c}{Mean} & \multicolumn{1}{c}{Median} & \multicolumn{1}{c}{Max} & \multicolumn{1}{c}{SD} & \multicolumn{1}{c}{best (\#)}&\multicolumn{1}{c|}{1st-2nd best (\#)}            & \multicolumn{1}{c}{Min} & \multicolumn{1}{c}{Mean} & \multicolumn{1}{c}{Median} & \multicolumn{1}{c}{Max} & \multicolumn{1}{c}{SD} & \multicolumn{1}{c}{best (\#)} & \multicolumn{1}{c}{1st-2nd best (\#)} \\
\hline
 
 FT1 & \textbf{0.014 }& 0.328 & 0.279 & 1.116 & 0.235 & 14 & \textit{31} & \textbf{0.005} & 0.163 & 0.146 & 0.517 & \textit{0.113} & 15 & \textit{ 29} \\
FT21 & \textbf{0.014 }& \textit{0.287} & \textit{0.250} & \textit{0.940} & \textit{0.200} & \textit{17} & \textit{37} & \textbf{0.005} & \textbf{0.143 }& \textbf{0.129} & \textbf{0.393} & \textbf{0.108} & \textit{19} & \textbf{44} \\
FT63 &\textbf{ 0.014} & \textbf{0.282} & \textbf{0.206} & 0.984 & 0.236 & \textbf{29} &\textbf{ 43 }& \textbf{0.005} &\textit{ 0.147} & \textit{0.141} & 0.683 & 0.118 & \textbf{23} & \textit{40} \\
G-EDF & \textbf{0.014 }& 0.430 & 0.367 & 1.337 & 0.300 & \textit{7} & 21 & \textbf{0.005} & 0.242 & 0.217 & 0.940 & 0.186 & {15} & 20 \\
G-N & \textbf{0.014} & 0.892 & 0.874 & 2.175 & 0.385 & \textbf{1} & \textbf{1} & \textbf{0.005} & 0.315 & 0.287 & 1.152 & 0.203 & {4} & {7} \\
G-t & \textbf{0.014} & 0.399 & 0.367 & 2.351 & \textit{0.322} & 16 & \textit{20} & \textit{0.012} & 0.274 & 0.235 & 1.540 & 0.225 & {4} & 18 \\
GAS & \textbf{0.014 }& 0.424 & 0.367 & 1.293 & 0.324 & 15 & 28 & \textbf{0.005} & 0.191 & 0.164 & 0.693 & 0.140 & 15 & 31 \\
Historical & \textit{0.030} & 0.348 & 0.323 & \textbf{0.852} & \textbf{0.172} & 10 & 25 & \textbf{0.005} & 0.220 & 0.199 & \textit{0.499} & 0.119 & 7 & 17 \\

    \hline
          & \multicolumn{7}{c|}{VaR (5\%)} & \multicolumn{7}{c}{VaR (10\%)} \\
          \hline
          & \multicolumn{1}{c}{Min} & \multicolumn{1}{c}{Mean} & \multicolumn{1}{c}{Median} & \multicolumn{1}{c}{Max} & \multicolumn{1}{c}{SD} & \multicolumn{1}{c}{best (\#) }& \multicolumn{1}{c|}{1st-2nd best (\#)}            & \multicolumn{1}{c}{Min} & \multicolumn{1}{c}{Mean} & \multicolumn{1}{c}{Median} & \multicolumn{1}{c}{Max} & \multicolumn{1}{c}{SD} & \multicolumn{1}{c}{best (\#)} & \multicolumn{1}{c}{1st-2nd best (\#)} \\

    \hline
FT1 & \textbf{0.004} & \textit{0.075} & \textbf{0.054} & 0.270 & \textit{0.058} & \textbf{19} & \textit{35} & \textit{0.004} & \textbf{0.049} & \textbf{0.045} & \textbf{0.133} & \textbf{0.030} & \textbf{22} & \textbf{39} \\
FT21 & \textbf{0.004} & \textbf{0.072} &\textit{ 0.065} & \textbf{0.217} & \textbf{0.049} & \textit{16} & \textbf{37} & \textbf{0.001} & 0.091 & 0.092 & 0.206 & 0.048 & 10 & 17 \\
FT63 & \textbf{0.004} & 0.124 & 0.114 & 0.374 & 0.087 & \textit{16} & 26 & 0.012 & 0.147 & 0.147 & 0.286 & 0.072 & \textbf{2} & 5 \\
G-EDF & \textbf{0.004} & 0.154 & 0.093 & 0.808 & 0.156 & 9 & 22 & \textit{0.004} & 0.100 & 0.074 & 0.451 & 0.099 & 10 & 18 \\
G-N & \textit{0.005} & 0.145 & 0.097 & 0.781 & 0.135 & 6 & 16 & 0.012 & 0.170 & 0.161 & 0.561 & 0.096 & 3 & 5 \\
G-t & \textbf{0.004} & 0.189 & 0.146 & 1.257 & 0.194 & 9 & 15 & \textit{0.004} & 0.127 & 0.077 & 0.940 & 0.144 & 9 & 18 \\
GAS & \textbf{0.004} & 0.090 & 0.071 & 0.314 & 0.073 & 15 & 28 & \textbf{0.001} & \textit{0.065} & \textit{0.056 }& \textit{0.198} & \textit{0.046} & \textit{16 }& \textit{35} \\
Historical & \textbf{0.004} & 0.120 & 0.120 & \textit{0.261} & 0.069 & 9 & 19 & \textit{0.004} & 0.070 & 0.065 & 0.226 & \textit{0.046} & \textit{18} & 28 \\
PT1 & & & & & & & & 0.005 & 0.170 & 0.173 & 0.389 & 0.074 & \textbf{1} & \textbf{3} \\
PT21 & & & & & & & & \textit{0.004} & 0.089 & 0.080 & 0.235 & 0.054 & 10 & 16 \\
PT63 & & & & & & & & 0.010 & 0.125 & 0.118 & 0.279 & 0.064 & 1 & 5 \\

\hline
  \end{tabular}}%
    \caption{Summary statistics of the $|1-AE|$ values over the out-of-sample period from January 2015 to September 2023 for the eleven models. Additionally, we report the count of stocks for which each of the considered models was the best (achieved lowest value of $|1-AE|$) or was within the top two models (1st-2nd best). In case of a tie, equal ranks were given. The values are highlighted using \textbf{bold} for the best values and \textit{italicized} for the second-best in each column.}
  \label{table_ae}%
\end{sidewaystable}%

\begin{sidewaystable}[htbp]
  \centering
\resizebox{1.0\textwidth}{!}{
  \begin{tabular}{l|cccccccc|ccccccccccc}
          & \multicolumn{8}{c|}{VaR (1\%)} & \multicolumn{11}{c}{VaR (2.5\%)}\\
          \hline
          & \multicolumn{1}{l}{FT1} & \multicolumn{1}{l}{FT21} & \multicolumn{1}{l}{FT63} & \multicolumn{1}{l}{G-EDF} & \multicolumn{1}{l}{G-N} & \multicolumn{1}{l}{G-t} & \multicolumn{1}{l}{GAS} & \multicolumn{1}{l|}{Historical} &        \multicolumn{1}{l}{FT1} & \multicolumn{1}{l}{FT21} & \multicolumn{1}{l}{FT63} & \multicolumn{1}{l}{G-EDF} & \multicolumn{1}{l}{G-N} & \multicolumn{1}{l}{G-t} & \multicolumn{1}{l}{GAS} & \multicolumn{1}{l}{Historical} & \multicolumn{1}{l}{PT1} & \multicolumn{1}{l}{PT21} & \multicolumn{1}{l}{PT63} \\
          \hline
    FT1   & 0.000 & 0.041 & 0.046 & -0.103 & -0.564 & -0.072 & -0.096 & -0.020 &        0.000 & 0.020 & 0.016 & -0.079 & -0.152 & -0.111 & -0.028 & -0.057 &       &       &  \\
         &	     & (0.105)&	(0.093)	&(0.995)&	(1)&	(0.956)&	(0.989)&	(0.75)&	 	&(0.108)	&(0.177)&	(1)&	(1)&	(1)&	(0.93)&	(0.999)&&&\\
    FT21  & -0.041 & 0.000 & 0.005 & -0.143 & -0.605 & -0.112 & -0.137 & -0.061 &        -0.020 & 0.000 & -0.004 & -0.099 & -0.172 & -0.131 & -0.048 & -0.077 &       &       &  \\
&(0.895)&	 	&(0.434)&	(1)	&(1)	&(0.997)&	(1)&	(0.986)&	(0.892)&	 &	(0.605)&	(1)	&(1)&	(1)&	(0.995)&	(1)&&&\\
    FT63  & -0.046 & -0.005 & 0.000 & -0.149 & -0.610 & -0.117 & -0.142 & -0.066 &        -0.016 & 0.004 & 0.000 & -0.095 & -0.167 & -0.126 & -0.043 & -0.072 &       &       &  \\
&(0.907)&	(0.566)&	 &	(1)&	(1)	&(0.997)&	(1)&	(0.985)&	(0.823)	&(0.395)&	& 	(1)	&(1)&	(1)	&(0.988)&	(1)&&&\\
    G-EDF & 0.103***& 0.143*** & 0.149*** & 0.000 & -0.462 & 0.031 & 0.007 & 0.082** &        0.079*** & 0.099*** & 0.095*** & 0.000 & -0.072 & -0.031 & 0.052** & 0.023 &       &       &  \\
&(0.005)&	(0)	&(0)	& 	&(1)&	(0.249)&	(0.442)&	(0.012)&	(0)&	(0)	&(0)&	 &	(0.994)&	(0.848)&	(0.017)	&(0.164)&&&\\
    G-N   & 0.564*** & 0.605*** & 0.610*** & 0.462*** & 0.000 & 0.493*** & 0.468*** & 0.544*** &        0.152*** & 0.172*** & 0.167*** & 0.072*** & 0.000 & 0.041 & 0.124*** & 0.095*** &       &       &  \\
&(0)&	(0)&	(0)	&(0)&	& 	(0)&	(0)	&(0)&	(0)&	(0)	&(0)&	(0.006)&	& 	(0.098)&	(0)	&(0)&&&\\
    G-t   & 0.072* & 0.112*** & 0.117*** & -0.031 & -0.493 & 0.000 & -0.025 & 0.051        & 0.111*** & 0.131*** & 0.126*** & 0.031 & -0.041 & 0.000 & 0.083*** & 0.054** &       &       &  \\
&(0.044)&	(0.003)&	(0.003)&	(0.751)&	(1)&	 &	(0.696)&	(0.091)&	(0)	&(0)&	(0)&	(0.152)&	(0.902)&&	 	(0.002)&	(0.022)&&&\\
    GAS   & 0.096** & 0.137*** & 0.142*** & -0.007 & -0.468 & 0.025 & 0.000 & 0.076** &        0.028 & 0.048*** & 0.043** & -0.052 & -0.124 & -0.083 & 0.000 & -0.029 &       &       &  \\
&(0.011)&	(0)	&(0)&	(0.558)&	(1)&	(0.304)	&& 	(0.025)&	(0.070)&	(0.005)&	(0.012)&	(0.983)&	(1)	&(0.998)&&	 	(0.934)&&&\\
    Historical & 0.020 & 0.061** & 0.066** & -0.082 & -0.544 & -0.051 & -0.076 & 0.000        & 0.057*** & 0.077*** & 0.072*** & -0.023 & -0.095 & -0.054 & 0.029 & 0.000 &       &       &  \\
 &  (0.250)	&(0.014)&	(0.015)&	(0.988)&	(1)	&(0.909)&	(0.975)&&	 	(0.001)&	(0)&	(0)	&(0.836)	&(1)	&(0.978)&	(0.066)&&&&\\	 
          \hline
& \multicolumn{8}{c|}{VaR (5\%)} & \multicolumn{11}{c}{VaR (10\%)}\\
\hline
FT1   & 0.000 & 0.004 & -0.049 & -0.078 & -0.069 & -0.113 & -0.014 & -0.045 &        0.000 & -0.042 & -0.098 & -0.051 & -0.121 & -0.078 & -0.016 & -0.021 & -0.040 & -0.121 & -0.075 \\
&&(0.327)&	(1)&	(1)&	(1)	&(1)&	(0.929)&	(1)&	& 	(1)&	(1)&	(1)&	(1)&	(1)&	(0.997)&	(1)&	(1)&	(1)	&(1)\\\
    FT21  & -0.004 & 0.000 & -0.053 & -0.082 & -0.073 & -0.117 & -0.018 & -0.048 &        0.042*** & 0.000 & -0.056 & -0.009 & -0.078 & -0.036 & 0.026*** & 0.021*** & 0.003 & -0.079 & -0.033 \\
&(0.673)&	& 	(1)&	(1)&	(1)&	(1)&	(0.974)&	(1)&	(0)&	& 	(1)&	(0.774)&	(1)&	(0.987)&	(0)&	(0.001)&	(1)&	(0.368)&	(1)\\
    FT63  & 0.049*** & 0.053*** & 0.000 & -0.029 & -0.020 & -0.064 & 0.035*** & 0.004 &        0.098*** & 0.056*** & 0.000 & 0.047*** & -0.022 & 0.020 & 0.082*** & 0.077*** & 0.059 & -0.023*** & 0.023** \\
&(0)	&(0)&	& 	(0.941)&	(0.886)&	(0.998)&	(0.002)&	(0.352)&	(0)	&(0)&	 &	(0)	&(0.963)	&(0.116)&	(0)&	(0)&	(0.983)	&(0)&	(0.012)\\
    G-EDF & 0.078*** & 0.082*** & 0.029 & 0.000 & 0.009 & -0.035 & 0.064*** & 0.034* &        0.051*** & 0.009 & -0.047 & 0.000 & -0.070 & -0.027 & 0.035*** & 0.030*** & 0.011 & -0.070 & -0.024 \\
 &(0)&	(0)&	(0.059)&	 	&(0.335)&	(0.91)&	(0)	&(0.03)	&(0)&	(0.226)&	(1)	& &	(1)	&(0.9310&	(0.001)&	(0.005)&	(1)	&(0.171)&	(0.976)\\
    G-N   & 0.069*** & 0.073*** & 0.020 & -0.009 & 0.000 & -0.044 & 0.055*** & 0.025 &        0.121*** & 0.078*** & 0.022* & 0.070*** & 0.000 & 0.043*** & 0.105*** & 0.100*** & 0.081 & -0.001*** & 0.045*** \\
&(0)&	(0)	&(0.114)&	(0.665)&&	 	(0.963)&	(0)	&(0.06)&	(0)	&(0)	&(0.037)&	(0)&	 &	(0.01)&	(0)	&(0)&	(0.516)&	(0)&	(0)\\
    G-t   & 0.113*** & 0.117*** & 0.064*** & 0.035 & 0.044* & 0.000 & 0.099*** & 0.069*** &        0.078*** & 0.036** & -0.020 & 0.027 & -0.043 & 0.000 & 0.062*** & 0.057*** & 0.038 & -0.043*** & 0.003 \\
&(0)&	(0)&	(0.002)&	(0.09)&	(0.037)&	& 	(0)	&(0.001)&	(0)&	(0.013)&	(0.884)&	(0.069)&	(0.99)&	& 	(0)	&(0)&	(0.994)&	(0.009)&	(0.432)\\
    GAS   & 0.014 & 0.018* & -0.035 & -0.064 & -0.055 & -0.099 & 0.000 & -0.030 &        0.016*** & -0.026 & -0.082 & -0.035 & -0.105 & -0.062 & 0.000 & -0.005 & -0.024 & -0.105 & -0.059 \\
&(0.071)	&(0.026)&	(0.998)&	(1)&	(1)	&(1)&	& 	(0.998)&	(0.003)&	(1)	&(1)&	(0.999)&	(1)	&(1)	& &	(0.769)&	(1)&	(0.999)	&(1)\\
    Historical & 0.045*** & 0.048*** & -0.004 & -0.034 & -0.025 & -0.069 & 0.030*** & 0.000 &        0.021*** & -0.021 & -0.077 & -0.030 & -0.100 & -0.057 & 0.005 & 0.000 & -0.019 & -0.100 & -0.054 \\
  &  (0)&	(0)	&(0.648)&	(0.97)&(0.94)&	(0.999)&	(0.002)&	& 	(0)&	(0.999)	&(1)&(0.995)&	(1)&	(1)&	(0.231)&	& 	(1)&	(0.994)	&(1)\\
    PT1   &       &       &       &       &       &       &       &       &        0.121*** & 0.079*** & 0.023** & 0.070*** & 0.001 & 0.043*** & 0.105*** & 0.100*** & 0.000& 0.082***  & 0.046*** \\
   &       &       &       &       &       &       &       &       &        (0)&	(0)&	(0.017)&	(0)&(0.484)&	(0.006)&	(0)&	(0)	& &	(0)&	(0) \\
    PT21  &       &       &       &       &       &       &       &       &        0.040*** & -0.003& -0.059 & -0.011 & -0.081 & -0.038 & 0.024*** & 0.019*** & -0.082& 0.000  & -0.036 \\
   &       &       &       &       &       &       &       &       &        (0)&	(0.632)&	(1)	&(0.829)	&(1)&	(0.991)&	(0.001)&	(0.006)	&(1)	& &	(1) \\
    PT63  &       &       &       &       &       &       &       &       &        0.075*** & 0.033*** & -0.023 & 0.024** & -0.045 & -0.003 & 0.059*** & 0.054*** & -0.046& 0.036***  & 0.000 \\
       &       &       &       &       &       &       &       &       &     (0)	&(0)	&(0.988)&	(0.024)&	(1)	&(0.568)&	(0)&	(0)	&(1)&	(0)&	\\ 
    \end{tabular}}
    \caption{We report the out-of-sample difference of $|1-AE|$, over the period from January 2015 to September 2023, of each model in the selected column from that of the benchmark in the selected row for value-at-risk at different levels. That is, we report the mean, across the stocks, of $|1-AE|_i-|1-AE|_j$ where $|1-AE|_i$ is the value from model in ith row and $j$ is the value from the model in jth column.  We adopt the following formatting: *, ** and *** denote whether the $t$-test of equal predictive accuracy is rejected at the 5\%, 2.5\% and 1\% level of significance, respectively. The hypothesis being tested is \( H_0 : 
|1-AE|_i = |1-AE|_j \) against a one-sided alternative \( H_1 : |1-AE|_i > |1-AE|_j \).}%
  \label{table_aetest}%
\end{sidewaystable}%

Tables \ref{table_uc}, \ref{table_cc} and \ref{table_dq} present the out-of-sample results of these tests, namely UC, CC and DQ respectively, for various confidence levels. The out-of-sample results are also depicted via bar plots in Figures \ref{ucplots}-\ref{dqplots} in Appendix I. For each $VaR_{\alpha}$, we present for how many assets we failed to reject the null hypothesis of correct model performance at various confidence levels. The higher the count, the better the model. From these tables, it is evident that the TimesFM shows promising results; in particular, the fine-tuned TimesFM model with one prediction length (FT1) either comfortably beats or performs fairly in comparison to the benchmark models across all the three tests. More specifically, for the UC test, the fine-tuned TimesFM models with various prediction lengths and the historical model showed the highest number of assets where the null hypothesis was not rejected, indicating that these models had the most consistent unconditional coverage. What's more remarkable is that the simple rolling method (historical) outperforms the GAS model in UC test. Again, in conditional coverage tests, the FT model with 1 prediction length performed significantly better than all other models. Finally, the performance of FT1 in DQ test is comparable to the benchmark models if not better. In conclusion,  the {FT$_{1}$} model demonstrates strong performance, providing robust competition to the GAS model across multiple tests, while TimesFM models with longer prediction lengths show significant weaknesses, particularly in capturing the conditional and dynamic aspects of the VaR process.

\begin{table}[htbp]
  \centering
   \resizebox{0.98\textwidth}{!}{ \begin{tabular}{rr|rrrrrrrrrrr}
\multicolumn{10}{c}{Unconditional coverage Test}\\
\hline
    \multicolumn{2}{l|}{Significance Level}        & \multicolumn{1}{l}{FT1} & \multicolumn{1}{l}{FT21} & \multicolumn{1}{l}{FT63} & \multicolumn{1}{l}{G-EDF} & \multicolumn{1}{l}{G-N} & \multicolumn{1}{l}{G-t} & \multicolumn{1}{l}{GAS} & \multicolumn{1}{l}{Historical}& \multicolumn{1}{l}{PT1} & \multicolumn{1}{l}{PT21} & \multicolumn{1}{l}{PT63} \\
\hline
&&\multicolumn{8}{c}{VaR (1\%)}\\
  & 99\%    & 77    & \textbf{83}    & \textit{80}    & 66    & 17    & 73    & 69    & \textbf{83} \\
          & 97.5\%  & 75    & \textbf{81}    & 75    & 65    & 15    & 67    & 65    & \textit{76} \\
          & 95\%    & \textit{65 }   & \textbf{68}    & \textbf{68}    & 55    & 11    & 56    & 56    & 64 \\
          \hline
&&\multicolumn{8}{c}{VaR (2.5\%)}\\
    & 99\%    & 85    & \textbf{88}    & \textit{87}    & 71    & 58    & 66    & 79    & 75 \\
          & 97.5\%  & \textit{83}    & 81    & \textbf{84}   & 65    & 51    & 61    & 76    & 67 \\
          & 95\%    & \textit{78}    &{77}    & \textbf{80}   & 59    & 41    & 51    & 69    & 59 \\
          \hline
&&\multicolumn{8}{c}{VaR (5\%)}\\
 & 99\%    & \textit{91}    & \textbf{92}    & 75    & 72    & 77    & 72    & 87    & 86 \\
& 97.5\%  & \textit{90}    & \textbf{91}    & 72    & 69    & 73    & 65    & 84    & 81 \\
& 95\%    & \textit{87}    & \textbf{88}    & 64    & 67    & 67    & 59    & 81    & 78 \\
          \hline
&&\multicolumn{8}{c}{VaR (10\%)}\\
& 99\%    & \textbf{92}    & 84    & 49     & 75    & 49    & 70    & 89    & \textit{{90}} & 40    & 85    & 63   \\
& 97.5\%    & \textbf{92}    & 78    & 43     & 72    & 36    & 65    & 85    & \textit{86 }& 34    & 74    & 56   \\
& 95\%  & \textbf{91}    & {66}    & 34      &70    & 31    & 61    & 81    & \textit{82} & 24    & 68    & 51  \\
   
    \end{tabular}}%
     \caption{This table shows the number of assets (out of 92) for which we failed to reject the null hypothesis in the Unconditional Coverage (UC) test, at various Value-at-Risk (VaR) levels, on the out-of-sample forecasts for the period January 2015 to September 2023. A higher number indicates better model performance, as it reflects a closer alignment between observed and expected VaR violations for each specified model. Results are reported for four VaR confidence levels (1\%, 2.5\%, 5\%, and 10\%) across different models. The best values are highlighted using \textbf{bold} and \textit{italicized} for the second-best in each row.}
  \label{table_uc}%
\end{table}%

\begin{table}[htbp]
  \centering
   \resizebox{0.98\textwidth}{!}{ \begin{tabular}{rr|rrrrrrrrrrr}
\multicolumn{10}{c}{Conditional Coverage Test}\\
\hline
    \multicolumn{2}{l|}{Significance Level}        & \multicolumn{1}{l}{FT1} & \multicolumn{1}{l}{FT21} & \multicolumn{1}{l}{FT63} & \multicolumn{1}{l}{G-EDF} & \multicolumn{1}{l}{G-N} & \multicolumn{1}{l}{G-t} & \multicolumn{1}{l}{GAS} & \multicolumn{1}{l}{Historical}& \multicolumn{1}{l}{PT1} & \multicolumn{1}{l}{PT21} & \multicolumn{1}{l}{PT63} \\
\hline
&&\multicolumn{8}{c}{VaR (1\%)}\\
& 99\%    & \textbf{79}    & 53    & 43    & 65    & 21    & \textit{72 }   & \textit{72 }   & 43 \\
& 97.5\%  & \textbf{70}  & 50    & 42    & 62    & 16    & 65    &\textit{ 68 }   & 43 \\
& 95\%    &\textbf{67}    & 47    & 38    & \textit{61}    & 12    & 57    & \textit{61}    & 37 \\
\hline
&&\multicolumn{8}{c}{VaR (2.5\%)}\\
& 99\%   & \textbf{83}    & 43    & 40    & 67    & 58    & 61    & \textit{81}    & 33 &&&\\
& 97.5\%  & \textbf{72}   & 33    & 27    & 57    & 46    & 57    & \textit{70}   & 19 &&&\\
& 95\%    & \textbf{66}    & 29    & 19    & 54    & 40    & 52    & \textit{63}    & 15 &&&\\
          \hline
&&\multicolumn{8}{c}{VaR (5\%)}\\
& 99\%    & \textbf{82}    & 41    & 33    & 65    & 63    & 63    & \textit{80}    & 24 &&&\\
& 97.5\%  & \textbf{75}    & 31    & 16    & 56    & 59    & 52    & \textit{72}    & 17 &&&\\
& 95\%    & \textit{62}    & 22    & 10    & 52    & 52    & 47    & \textbf{66}    & 10&&& \\
\hline
&&\multicolumn{8}{c}{VaR (10\%)}\\
& 99\%    & \textbf{88}    & 30    & 12       & 58    & 27    & 55    & \textit{84}    & 40 & 41    & 31    & 23 \\
& 97.5\%    & \textbf{82}    & 22    & 4       & 49    & 22    & 50    & \textbf{82}    & 29  & 32    & 19    & 17 \\
& 95\%  & \textbf{79}    & 14    & 3       & 47    & 16    & 46    & \textit{74}    & 16 & 28    & 16   & 12  \\ 

\end{tabular}}%
\caption{This table shows the number of assets (out of 92) for which we failed to reject the null hypothesis in the Conditional Coverage (CC) test, at various value-at-risk (VaR) levels, on the out-of-sample forecasts for the period January 2015 to September 2023. A higher number indicates better model performance, as it reflects a closer alignment between observed and expected VaR violations for each specified model. Results are reported for four VaR confidence levels (1\%, 2.5\%, 5\%, and 10\%) across different models. The best-performing values are highlighted in bold, while the second-best values are italicized.}

  \label{table_cc}%
\end{table}%

\begin{table}[htbp]
  \centering
   \resizebox{0.98\textwidth}{!}{ \begin{tabular}{rr|rrrrrrrrrrr}
\multicolumn{10}{c}{Dynamic Quantile (DQ) Test}\\
\hline
    \multicolumn{2}{l|}{Significance Level}        & \multicolumn{1}{l}{FT1} & \multicolumn{1}{l}{FT21} & \multicolumn{1}{l}{FT63} & \multicolumn{1}{l}{G-EDF} & \multicolumn{1}{l}{G-N} & \multicolumn{1}{l}{G-t} & \multicolumn{1}{l}{GAS} & \multicolumn{1}{l}{Historical}& \multicolumn{1}{l}{PT1} & \multicolumn{1}{l}{PT21} & \multicolumn{1}{l}{PT63} \\
\hline
&&\multicolumn{8}{c}{VaR (1\%)}\\
& 99\%    & \textit{53}   & 35    & 18    & 46    & 9     & 46    & \textbf{55}    & 8 &&&\\
& 97.5\%  & \textit{43}    & 31    & 14    & 40    & 6     & 42    & \textbf{47}    & 6 &&&\\
& 95\%    & \textit{40}    & 27    & 9     & 37    & 6     & 37    & \textbf{43}    & 3 &&&\\
\hline
&&\multicolumn{8}{c}{VaR (2.5\%)}\\
& 99\%    & \textbf{57}    & 17    & 5     & 51    & 43    & 49    & \textit{55}    & 1 &&&\\
& 97.5\%  & \textbf{50}    & 14    & 3     & 39    & 38    & 40    & \textit{46}    & 0 &&&\\
& 95\%    & \textbf{39}     & 7    & 2     & 36    & 30    & \textit{38}    & \textbf{39}    & 0 &&&\\
\hline
&&\multicolumn{8}{c}{VaR (5\%)}\\
& 99\%    & \textit{56}   & 17    & 3     & 51    & 52    & 46    & \textbf{65}    & 4 &&&\\
& 97.5\%  & \textit{46}     & 9    & 2     & 43    & \textit{46}    & 40    & \textbf{57}    & 1 &&&\\
& 95\%    & \textit{38}     & 6    & 0     & \textit{38}    & 41    & 34    & \textbf{49}    & 0 &&&\\
\hline
&&\multicolumn{8}{c}{VaR (10\%)}\\
& 99\%    & \textit{61}     & 9    & 0          & 44    & 27    & 42    & \textbf{73}    & 1 & 23     & 4   & 1\\
&  97.5\%     & \textit{44}     & 5    & 0        & 38    & 22    & 38    & \textbf{65}    & 0  & 14     & 2    & 1 \\
&    95\%   & \textit{35}     & 3    & 0         & 31    & 15    & 28    & \textbf{61}    & 0 & 10     & 2    & 1 \\
\end{tabular}}%
\caption{This table presents the number of assets (out of 92) that passed the Dynamic Quantile (DQ) test, at various value-at-risk (VaR) levels, on the out-of-sample forecasts for the period January 2015 to September 2023, across different models and significance levels. A higher count indicates better model performance. The table reports results for four VaR confidence levels (1\%, 2.5\%, 5\%, and 10\%). The best-performing values are bolded, and the second-best values are italicized within each row to highlight relative model effectiveness.}
  \label{table_dq}%
\end{table}%

\subsection{Quantile Loss}

Table \ref{table_qloss} presents the summary statistics of quantile loss scores across 92 stocks for out of sample data from January 2015 to September 2023. We observe that, at 1\% VaR, the mean and median quantile score is consistent across the models, with FT1 and FT21 exhibiting lower mean values, 0.062 and 0.065, respectively. This indicates a better performance of fine-tuned TimesFM models compared to GARCH models (G-EDF, G-t, G-N) that have higher mean values and higher values for maximum quantile score. Further, a higher value of the standard deviation for GARCH models shows that these models have high variability in their VaR forecasts which, in turn, reflects that these models are less stable in extreme market scenarios. These observations remain valid at 2.5\% VaR forecasts, with FT1 and FT21 showing lower mean and median quantile scores, while GARCH models still show higher values for maximum quantile score and the standard deviation. For VaR at 5\% and  10\% levels, the historical model performs similar to fine-tuned TimesFM model whereas GARCH models continue to show higher maximum quantile score and greater variability, thus making them less robust and more sensitive to extreme market conditions. In addition, GAS models obtain lower values of mean and median quantile scores, similar to those obtained by FT models across all quantile levels. FT models are thus competitive to the benchmark models, in particular the GAS model which shows slightly higher values of maximum quantile loss and higher standard deviation, particularly at VaR 5\% and 10\% levels.

\begin{table}[htbp]
  \centering
    \resizebox{.98\textwidth}{!}{\begin{tabular}{l|ccccc|ccccc}
          & \multicolumn{5}{c}{VaR (1\%)}     & \multicolumn{5}{c}{VaR (2.5\%)}   \\
          \hline
          & \multicolumn{1}{c}{Min} & \multicolumn{1}{c}{Mean} & \multicolumn{1}{c}{Median} & \multicolumn{1}{c}{Max} & \multicolumn{1}{c|}{SD} & \multicolumn{1}{c}{Min} & \multicolumn{1}{c}{Mean} & \multicolumn{1}{c}{Median} & \multicolumn{1}{c}{Max} & \multicolumn{1}{c}{SD} \\
\hline

   FT1   & 0.036 & \textbf{0.062} & 0.061 & \textbf{0.097} & \textbf{0.013} & 0.071 & \textbf{0.118} & 0.116 & \textbf{0.191} & \textbf{0.024} \\
  
 FT21  & 0.039 & 0.065 & 0.065 & 0.101 & 0.014 & 0.074 & 0.123 & 0.123 & 0.195 & 0.025 \\
   FT63  & 0.04  & 0.069 & 0.069 & 0.108 & 0.015 & 0.077 & 0.128 & 0.127 & 0.209 & 0.026 \\
    G-EDF & 0.034 & 0.068 & 0.063 & 0.146 & 0.022 & \textbf{0.068} & 0.134 & 0.123 & 0.316 & 0.048 \\
    G-N   & 0.035 & 0.07  & 0.065 & 0.151 & 0.023 & 0.069 & 0.134 & 0.123 & 0.32  & 0.048 \\
    G-t   & \textbf{0.033} & 0.068 & 0.062 & 0.144 & 0.022 &\textbf{ 0.068} & 0.134 & 0.123 & 0.316 & 0.048 \\
    GAS   & 0.036 & \textbf{0.062} & \textbf{0.06}  & 0.103 & 0.014 & 0.07  & \textbf{0.118} & \textbf{0.115} & 0.198 & 0.025 \\
    Historical & 0.041 & 0.068 & 0.067 & 0.115 & {0.015} & 0.076 & 0.127 & 0.127 & {0.219} & 0.027 \\
          &       &       &       &       &       &       &       &       &       &  \\
          \hline
          & \multicolumn{5}{c}{VaR (5\%)}& \multicolumn{5}{c}{VaR (10\%)} \\
          \hline
          & \multicolumn{1}{c}{Min} & \multicolumn{1}{c}{Mean} & \multicolumn{1}{c}{Median} & \multicolumn{1}{c}{Max} & \multicolumn{1}{c|}{SD} & \multicolumn{1}{c}{Min} & \multicolumn{1}{c}{Mean} & \multicolumn{1}{c}{Median} & \multicolumn{1}{c}{Max} & \multicolumn{1}{c}{SD} \\
          \hline
    FT1   & 0.113 & 0.189 & 0.189 & 0.315 & 0.039 & 0.175 & 0.297 & 0.297 & \textbf{0.508} & \textbf{0.063} \\
     FT21  & 0.118 & 0.196 & 0.195 & 0.322 & 0.041 & 0.183 & 0.303 & 0.302 & 0.513 & 0.064 \\
   
    FT63  & 0.122 & 0.202 & 0.202 & 0.335 & 0.042 & 0.189 & 0.311 & 0.311 & 0.525 & 0.066 \\
    G-EDF & \textbf{0.111} & 0.221 & 0.199 & 0.556 & 0.084 & 0.175 & 0.352 & 0.312 & 0.931 & 0.143 \\
    G-N   & \textbf{0.111}& 0.221 & 0.199 & 0.553 & 0.084 & 0.175 & 0.353 & 0.315 & 0.92  & 0.141 \\
    G-t   &\textbf{0.111} & 0.221 & 0.199 & 0.556 & 0.085 & 0.175 & 0.352 & 0.312 & 0.931 & 0.144 \\
    GAS   & 0.112 & \textbf{0.188} & \textbf{0.186} &\textbf{0.32}  & \textbf{0.039} & \textbf{0.171} & \textbf{0.294} & \textbf{0.292} & 0.51  & \textbf{0.063} \\
    Historical & 0.12  & 0.201 & 0.202 & 0.356 & 0.043 & 0.187 & 0.308 & 0.311 & 0.536 & 0.067 \\
    PT1   &       &       &       &       &       & 0.178 & 0.3   & 0.299 & 0.512 & 0.064 \\
    PT21  &       &       &       &       &       & 0.185 & 0.306 & 0.305 & 0.524 & 0.065 \\
  
    PT63  &       &       &       &       &       & 0.189 & 0.31  & 0.31  & 0.537 & 0.067 \\
    \end{tabular}}%
\caption{This table presents summary statistics of the average quantile score for eleven VaR forecasting models, calculated over an out-of-sample period from January 2015 to September 2023. The quantile score measures the accuracy of VaR forecasts by penalizing deviations from the true quantile level, where lower scores indicate better model performance. Summary statistics include the minimum (Min), mean, median, maximum (Max), and standard deviation (SD) of the quantile score distribution for each model across 92 assets. Results are reported for four VaR confidence levels: 1\%, 2.5\%, 5\%, and 10\%. The lowest values in each column are bolded to highlight the best-performing model for each statistic and quantile level.}

  \label{table_qloss}%
\end{table}%

As the quantile scores are not easy to interpret, we calculate skill scores defined as the ratio of each model’s score to the score of a benchmark model for an individual asset, and then, we take the mean across assets to summarize the performance of models across multiple assets. More specifically, we report in Table \ref{table_dmtest}, the out-of-sample total quantile loss of each model relative to the benchmark model in the selected row. For instance, for VaR (1\%), the first row contains the cross-sectional average of the ratio of out-of-sample total quantile loss of every model relative to the out-of-sample total quantile loss of model FT21. In addition to the comparison of the total quantile loss, we also perform the Diebold–Mariano test for each pair of models where we test 
\begin{equation*}
H_0 : \text{total quantile loss}_i = \text{total quantile loss}_j, 
\end{equation*}
against a one-sided alternative 
\begin{equation*}
H_1 : \text{total quantile loss}_i > \text{total quantile loss}_j, 
\end{equation*}
where \( i \) denotes the model in the selected row, whereas \( j \) denotes the model in the selected column. Further, the formatting in the Tables is as follows: *, **, and *** denote whether the Diebold–Mariano (DM) test of equal predictive accuracy is rejected for more than 50\% of the assets at the 5\%, 2.5\% and 1\% level of significance.

\begin{table}[htbp]
    
  \resizebox{.78\textwidth}{!}{
  \begin{tabular}{l|ccccccccccc}
  \hline
          \multicolumn{11}{c}{VaR (1\%)} \\
          \hline
          & FT1 & FT21 & FT63 & G-EDF & G-N & G-t & GAS & Historical & PT1 & PT21& PT63\\
          \hline
FT1   & 0.000 & 1.061 & 1.123 & 1.097 & 1.131 & 1.095 & 1.005 & 1.109 && \\
FT21  & 0.944 & 0.000 & 1.058 & 1.034 & 1.066 & 1.032 & 0.949 & 1.044 &&\\
FT63  & 0.893 & 0.946 & 0.000 & 0.978 & 1.009 & 0.976 & 0.898 & 0.988 &&\\
G-EDF & 0.950 & 1.008 & 1.066 & 0.000 & 1.030 & 0.971** & 0.955 & 1.052 &&\\
G-N   & 0.923 & 0.979 & 1.036 & 0.971*** & 0.000 & 0.969*** & 0.928 & 1.022 &&\\
G-t   & 0.952 & 1.011 & 1.069 & 1.002 & 1.033 & 0.000 & 0.957 & 1.055&& \\
GAS   & 0.997 & 1.058 & 1.120 & 1.094 & 1.128 & 1.091 & 0.000 & 1.105&& \\
Historical & 0.905 & 0.959 & 1.014 & 0.991 & 1.021 & 0.988 & 0.909 & 0.000 &&\\
          \hline
          \multicolumn{11}{c}{VaR (2.5\%)} \\
          \hline
FT1   & 0.000 & 1.045 & 1.090 & 1.138 & 1.142 & 1.141 & 0.998 & 1.078&& \\
FT21  & 0.957 & 0.000 & 1.042 & 1.088 & 1.092 & 1.090 & 0.955 & 1.031 &&\\
FT63  & 0.919 & 0.960 & 0.000 & 1.044 & 1.048 & 1.046 & 0.917 & 0.989 &&\\
G-EDF & 0.929 & 0.970 & 1.011 & 0.000 & 1.003 & 1.002 & 0.927 & 0.999 &&\\
G-N   & 0.926 & 0.968 & 1.008 & 0.997*** & 0.000 & 0.998*** & 0.925 & 0.997&& \\
G-t   & 0.928 & 0.969 & 1.010 & 0.999*** & 1.002 & 0.000 & 0.926 & 0.998 &&\\
GAS   & 1.003 & 1.048 & 1.093 & 1.142 & 1.146 & 1.144 & 0.000 & 1.080 &&\\
Historical & 0.929* & 0.971 & 1.012 & 1.055 & 1.059 & 1.057 & 0.927* & 0.000&& \\
          \hline
          \multicolumn{11}{c}{VaR (5\%)} \\
          \hline
FT1   & 0.000 & 1.034 & 1.068 & 1.167 & 1.167 & 1.170 & 0.993 & 1.060 &&\\
FT21  & 0.967 & 0.000 & 1.033 & 1.128 & 1.128* & 1.130 & 0.960 & 1.025 &&\\
FT63  & 0.937 & 0.968 & 0.000 & 1.092 & 1.091 & 1.094 & 0.930 & 0.993 &&\\
G-EDF & 0.915 & 0.946 & 0.977 & 0.000 & 1.000*** & 1.001 & 0.908 & 0.969&& \\
G-N   & 0.914 & 0.945 & 0.976 & 0.999 & 0.000 & 1.001 & 0.908 & 0.968 &&\\
G-t   & 0.914 & 0.945 & 0.976 & 0.999*** & 0.999*** & 0.000 & 0.908 & 0.968&& \\
GAS   & 1.008 & 1.042 & 1.076 & 1.177 & 1.176 & 1.179 & 0.000 & 1.068 &&\\
Historical & 0.944*** & 0.976*** & 1.008** & 1.099** & 1.098** & 1.101* & 0.937*** & 0.000 &&\\
          \hline
          \multicolumn{11}{c}{VaR (10\%)} \\
          \hline
FT1   & 0.000 & 1.021 & 1.046 & 1.188 & 1.193** & 1.191 & 0.992* & 1.037 & 1.009 & 1.029 & 1.043 \\
FT21  & 0.980 & 0.000 & 1.025 & 1.163 & 1.167** & 1.165 & 0.971* & 1.016 & 0.988 & 1.008 & 1.022 \\
FT63  & 0.956 & 0.976 & 0.000 & 1.135 & 1.139 & 1.137 & 0.948 & 0.991 & 0.965 & 0.984 & 0.997 \\
G-EDF & 0.908 & 0.927 & 0.949 & 0.000 & 1.005*** & 1.001 & 0.900 & 0.940 & 0.916 & 0.934 & 0.946 \\
G-N   & 0.902 & 0.921 & 0.943 & 0.995 & 0.000 & 0.996 & 0.894 & 0.934 & 0.910 & 0.928 & 0.940 \\
G-t   & 0.907 & 0.926 & 0.949 & 0.999*** & 1.004*** & 0.000 & 0.899* & 0.940 & 0.915 & 0.933 & 0.946 \\
GAS   & 1.009 & 1.030 & 1.055 & 1.198 & 1.203** & 1.200 & 0.000 & 1.046 & 0.973 & 0.992 & 1.006 \\
Historical & 0.965*** & 0.985*** & 1.009*** & 1.144*** & 1.148*** & 1.146*** & 0.957*** & 0.000 & 0.973* & 0.992 & 1.006 \\
PT1   & 0.991*** & 1.012* & 1.037 & 1.177*** & 1.182*** & 1.180*** & 0.983*** & 1.028 & 0.000 & 1.020 & 1.034 \\
PT21  & 0.972*** & 0.992*** & 1.017*** & 1.154*** & 1.159*** & 1.156*** & 0.964*** & 1.008 & 0.981 & 0.000 & 1.014 \\
PT63  & 0.959 & 0.979 & 1.003 & 1.138 & 1.142* & 1.140 & 0.951* & 0.994 & 0.967 & 0.986 & 0.000 \\
    \end{tabular}}
    \caption{We report the out-of-sample total quantile loss, on the forecasts for the period January 2015 to September 2023, of each model in the selected column relative to the benchmark in the selected row for value-at-risk (1\%). Each number is a mean of such pairwise relative total quantile losses for each stock. we adopt the following formatting: *, ** and *** denotes whether the Diebold–Mariano (DM) test of equal predictive accuracy is rejected for more than 50\% of the assets at the 5\%, 2.5\% and 1\% level of significance, respectively, across individual tests for each asset. The hypothesis being tested is \( H_0 : \text{total quantile loss}_i = \text{total quantile loss}_j \) against a one-sided alternative \( H_1 : \text{total quantile loss}_i > \text{total quantile loss}_j \), where \( i \) denotes the model in the selected row, whereas \( j \) denotes the model in the selected column.}%
  \label{table_dmtest}

\end{table}

The cross-sectional mean total quantile score for the fine-tuned model TimesFM, FT1, is below one compared to other TimesFM models, such as FT21 and FT63. However, the results from the DM test do not show a preference for any model over the others. Further, we observe that the mean quantile score decreases as the prediction horizon increases, which suggests that TimesFM is better suited for short-horizon forecasts. The cross-sectional mean total quantile score for the TimesFM fine-tuned models is consistently below one, except for FT63, which scores 1.012 when TimesFM models are evaluated against the historical method as a benchmark. Moreover, the Diebold–Mariano test is rejected at the 5\%  or 10\%  significance level for FT1 and FT63 models, indicating that these models perform superior to the historical simulation method. Perhaps the most notable outcome is the competitive performance of the fine-tuned TimesFM models, particularly FT1, against well-established benchmark models such as GAS or Garch-EDF, which are well known for their VaR forecasting. Note that the mean quantile score for the GAS model is 1.005 when compared to FT1 as a benchmark. The conclusions remain valid when we compare the performance of these models in forecasting VaR at 2.5\% levels. Interestingly, the GAS model is excellent and challenging to beat in forecasting VaR.

We now focus on VaR forecasting at 5\% and 10\% levels, although not very important from the regulator's perspective, as these capture more common and less severe fluctuations in asset returns and, thus, are less likely to cause significant losses. Nonetheless, we report the comparison. It shows that the TimesFM model fails to beat the historical simulation method at almost all significant levels. What is more surprising compared with extant literature is that the GAS model fails to outperform the Historical Simulation method. A possible reconciliation of this is that the results naturally depend on the data period, split between training and test set, and finally, the results in the literature are usually compared for the stock indices, in contrast to our experiments on multiple stocks constituting the index. Observations for VaR (5\%) remain valid in Table for VaR (10\%). An appealing but not surprising observation that can be made from results for VaR (10\%) is that fine-tuning improves the model performance over the pre-trained model.

\section{Conclusion and discussion}\label{sec:conclusion_discussion}

Determining whether pre-trained foundation time-series models can outperform traditional econometric methods is a key question, as their success would mean that anyone with minimal statistical and mathematical knowledge could achieve relatively good VaR forecasts. In this paper, we evaluated the performance of Google’s foundation model for time series, TimesFM, in estimating Value-at-Risk (VaR) for the S\&P 100 index and its 91 constituents. We initially used the pre-trained TimesFM model without any task-specific training that yielded suboptimal results. However, fine-tuning the model with domain-specific data led to significant statistical improvements: In terms of Actual vs. Expected violations (AE), fine-tuned TimesFM outperforms all econometric benchmarks, including the GAS model, which is widely regarded as the leading method in the recent econometric literature \citep{patton2019dynamic}. Traditional models like GARCH (with different residual distributions) and empirical quantile estimates fall notably behind. These results remain consistent across different VaR levels, underscoring the robustness of TimesFM. Additionally, we assessed performance using the Quantile Score Loss Function, where TimesFM showed results comparable to the best econometric models.

As the results favor the foundation AI model, financial time-series modeling could shift toward data-driven approaches, partially reducing the need for mathematical modeling. Such a shift would have significant implications for both academia and the financial industry, as foundation models can be utilized with relatively minimal methodological expertise in econometrics. However, the adoption of foundation models is accompanied by notable challenges. A key challenge with these models is their “black-box” nature, which can make it difficult to understand how they arrive at their predictions. This is particularly concerning because we still have only a partial understanding of how these models operate, where they might fail, and what they are truly capable of, mainly due to their complex, emergent behaviors. While time series foundation models show promise in adapting to out-of-distribution scenarios, similar to how humans use ``common sense'' reasoning, they are also prone to hallucinations and may produce decisions that appear reasonable but are incorrect \citep{chakraborty2024hallucination}. This lack of transparency is especially problematic in high-stakes financial settings, where understanding the reasoning behind predictions is essential for effective risk management and decision-making. Moreover, it can create hurdles for regulatory acceptance, as regulators typically require clear and explainable models to meet compliance standards. Therefore, improving the interpretability of foundation models is critical to align with regulatory expectations and support their wider use in practice. 

Overall, our results---which are highly generalizable, having been conducted across over 90 assets with multiple settings---favor the pre-trained foundation model, highlighting its potential to enhance Value-at-Risk (VaR) forecasting by offering a completely new alternative to traditional econometric methods. We anticipate future research addressing various econometric and financial market challenges, such as portfolio optimization and volatility forecasting, as well as applications in other fields where time series forecasting methods are required. These include forecasting electricity and commodity prices \cite{bonato2024forecasting}, macroeconomic variables \cite{vcapek2023macroeconomic}, and service demand in hospitals \cite{tuominen2024forecasting}. We also emphasize the importance of critically examining the potential risks involved in applying foundation models to econometric problems.

\section*{Declaration of Generative AI and AI-assisted technologies in the writing process}

While preparing this work, the authors used GPT-4 to refine the language and enhance the clarity of the manuscript. After using this tool, the authors reviewed and edited the content as needed and take full responsibility for the content of the publication.


\section*{Acknowledgements} The first author acknowledges support from the European Union's Horizon Europe programme under the Marie Skłodowska-Curie Actions (Grant Agreement No. 101150609, Project: HiddenTipChains).

\section*{Conflicts of Interest}

The authors declare no conflicts of interest.

\bibliography{mybib}


%
%
%

\newpage
\section*{Appendix I}
\begin{figure}[ht!]
\begin{center}
\includegraphics[scale=0.30]{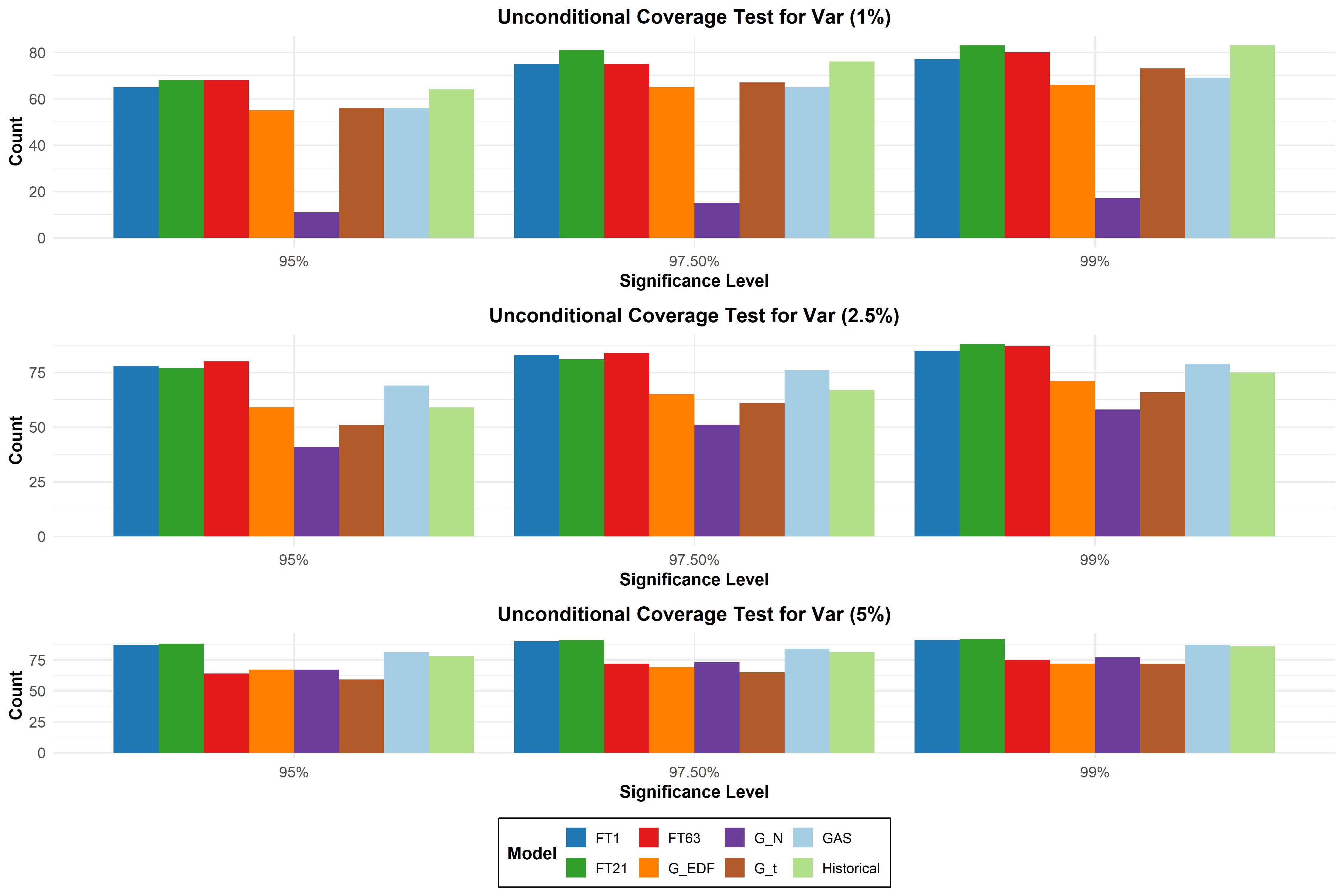}
\quad
\includegraphics[scale=0.250]{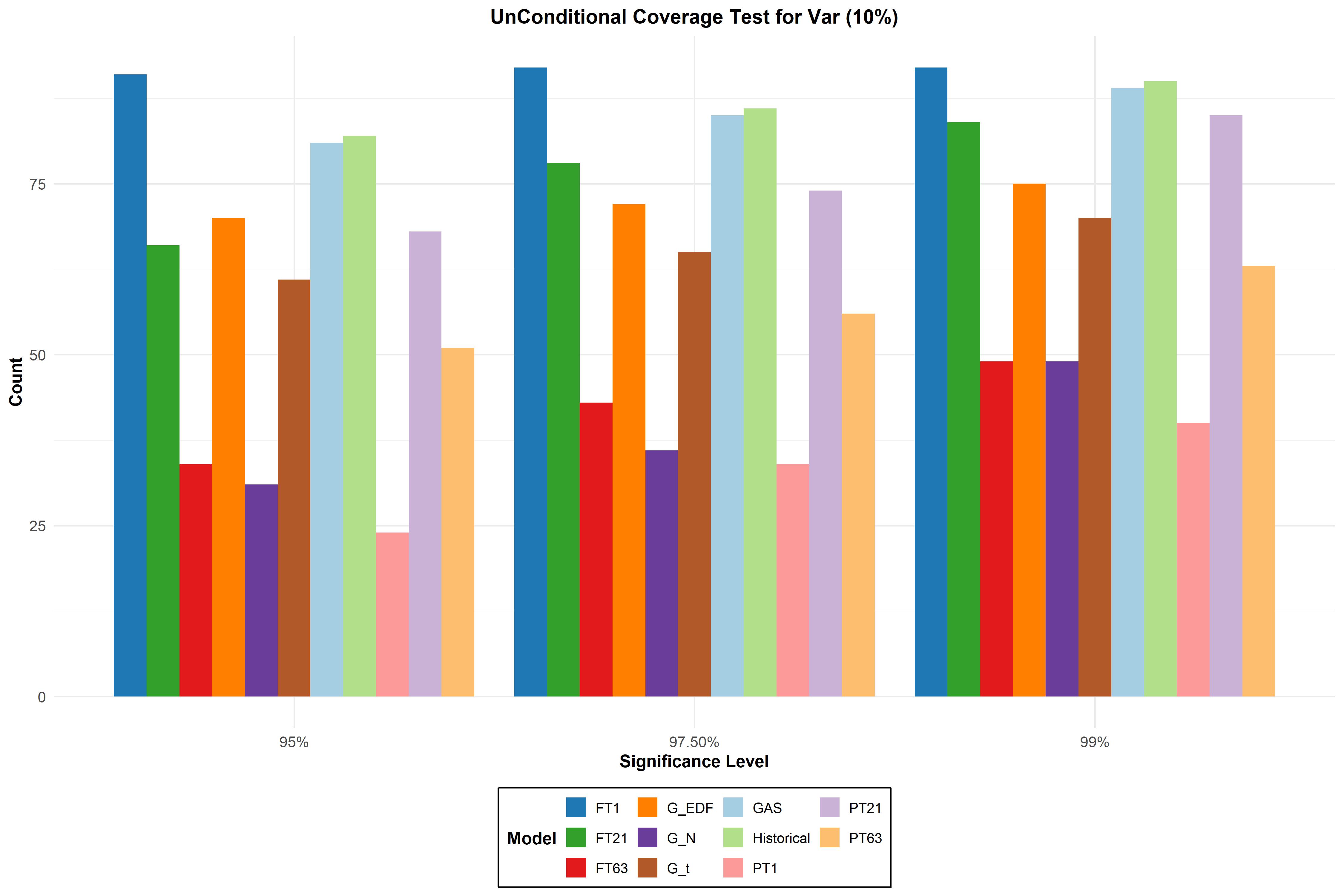}
\caption{The bar plots of number of assets (out of 92) for which we failed to reject the null hypothesis in the Unconditional Coverage (UC) test, at various value-at-risk (VaR) levels, on out-of-sample forecasts for the period January 2015 to September 2023. A higher number indicates better model performance, as it reflects a closer alignment between observed and expected VaR violations for each specified model. Results are reported for four VaR confidence levels (1\%, 2.5\%, 5\%, and 10\%) across different models.}
\label{ucplots}
\end{center}
\end{figure}

\begin{figure}[ht!]
\begin{center}
\includegraphics[scale=0.32]{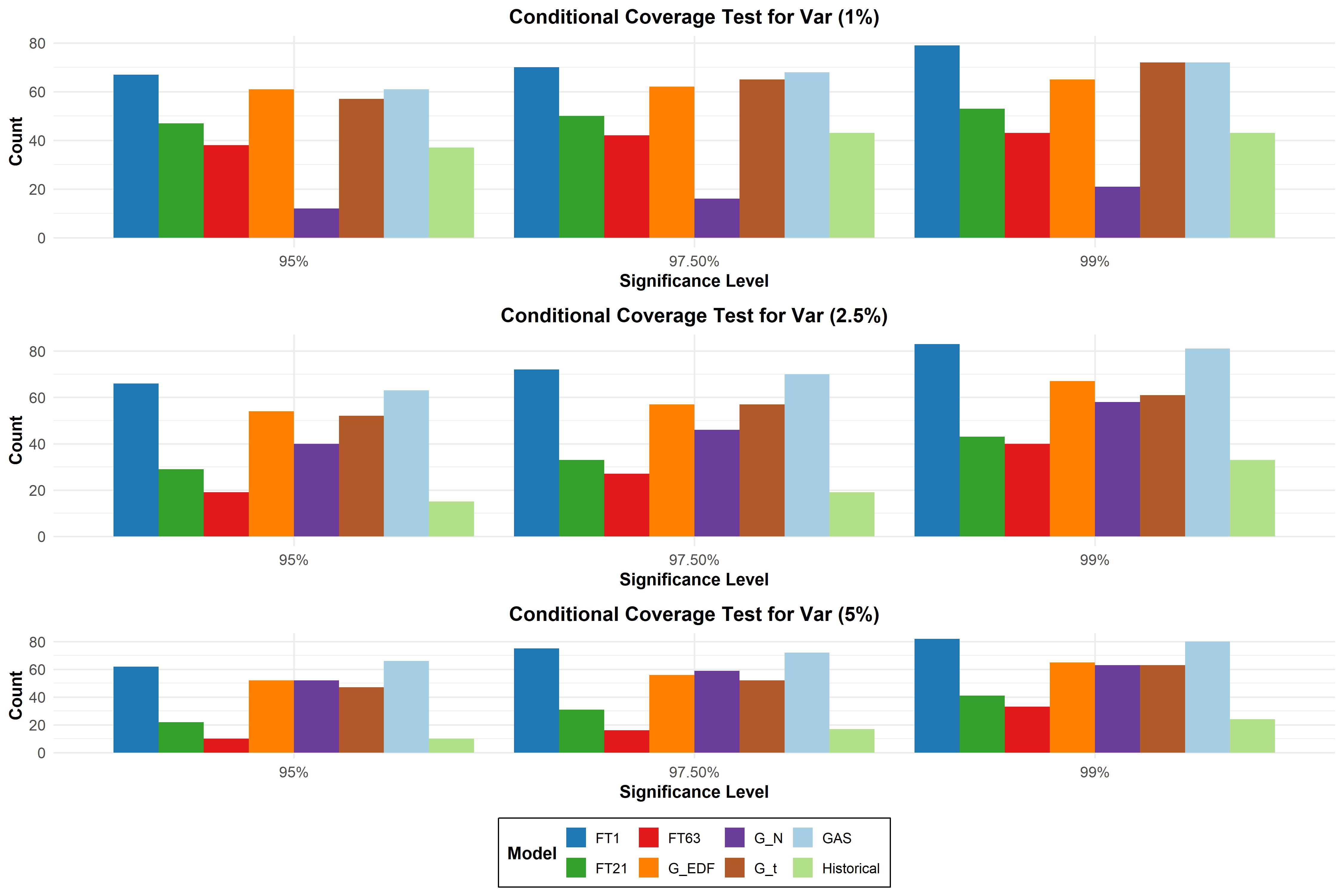}
\quad
\includegraphics[scale=0.30]{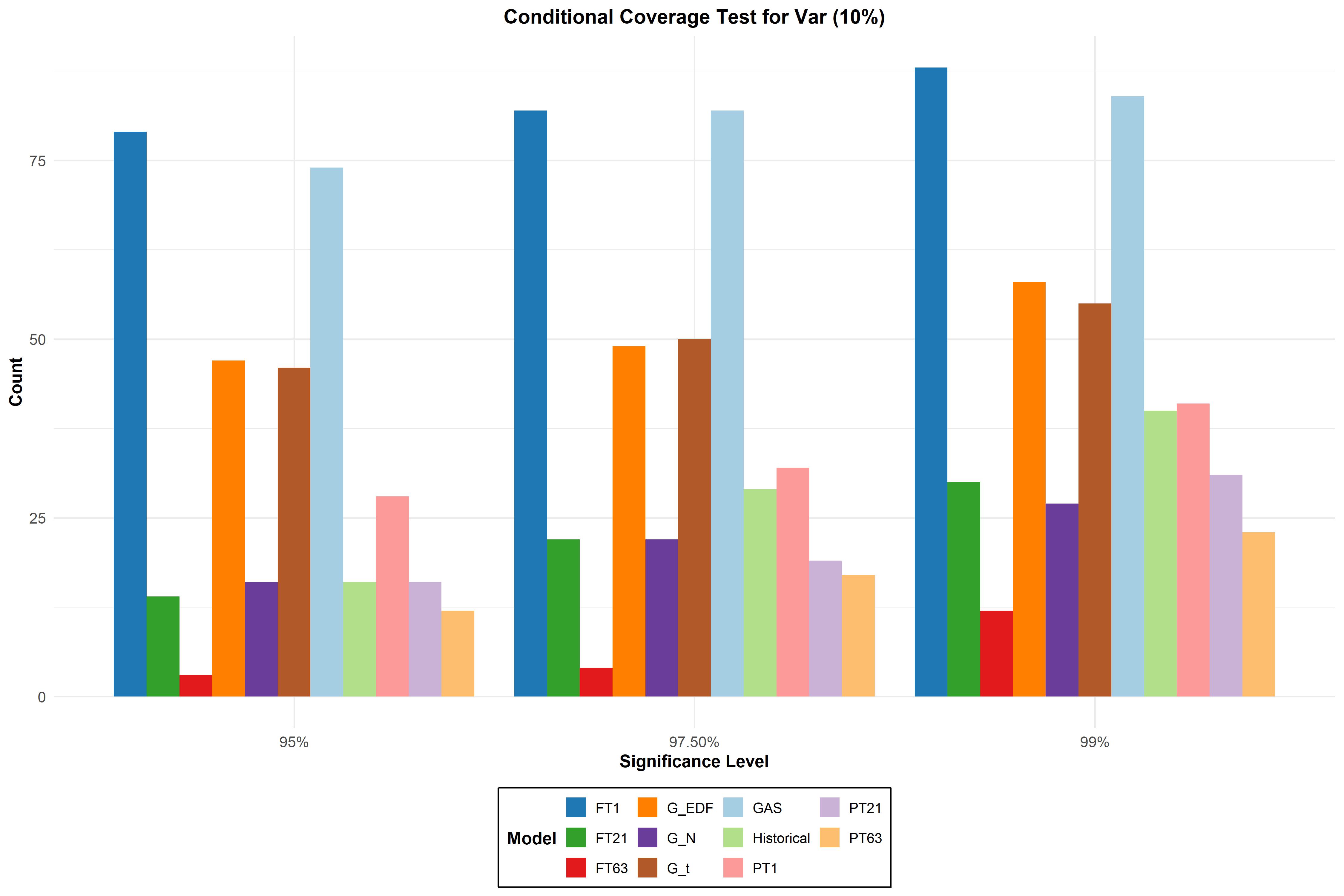}
\caption{The bar plot of the number of assets (out of 92) for which we failed to reject the null hypothesis in the Conditional Coverage (CC) test, at various value-at-risk (VaR) levels, on out-of-sample forecasts for the period January 2015 to September 2023. A higher number indicates better model performance, as it reflects a closer alignment between observed and expected VaR violations for each specified model. Results are reported for four VaR confidence levels (1\%, 2.5\%, 5\%, and 10\%) across different models.}
\label{ccplots}
\end{center}
\end{figure}

\begin{figure}[ht!]
\begin{center}
\includegraphics[scale=0.32]{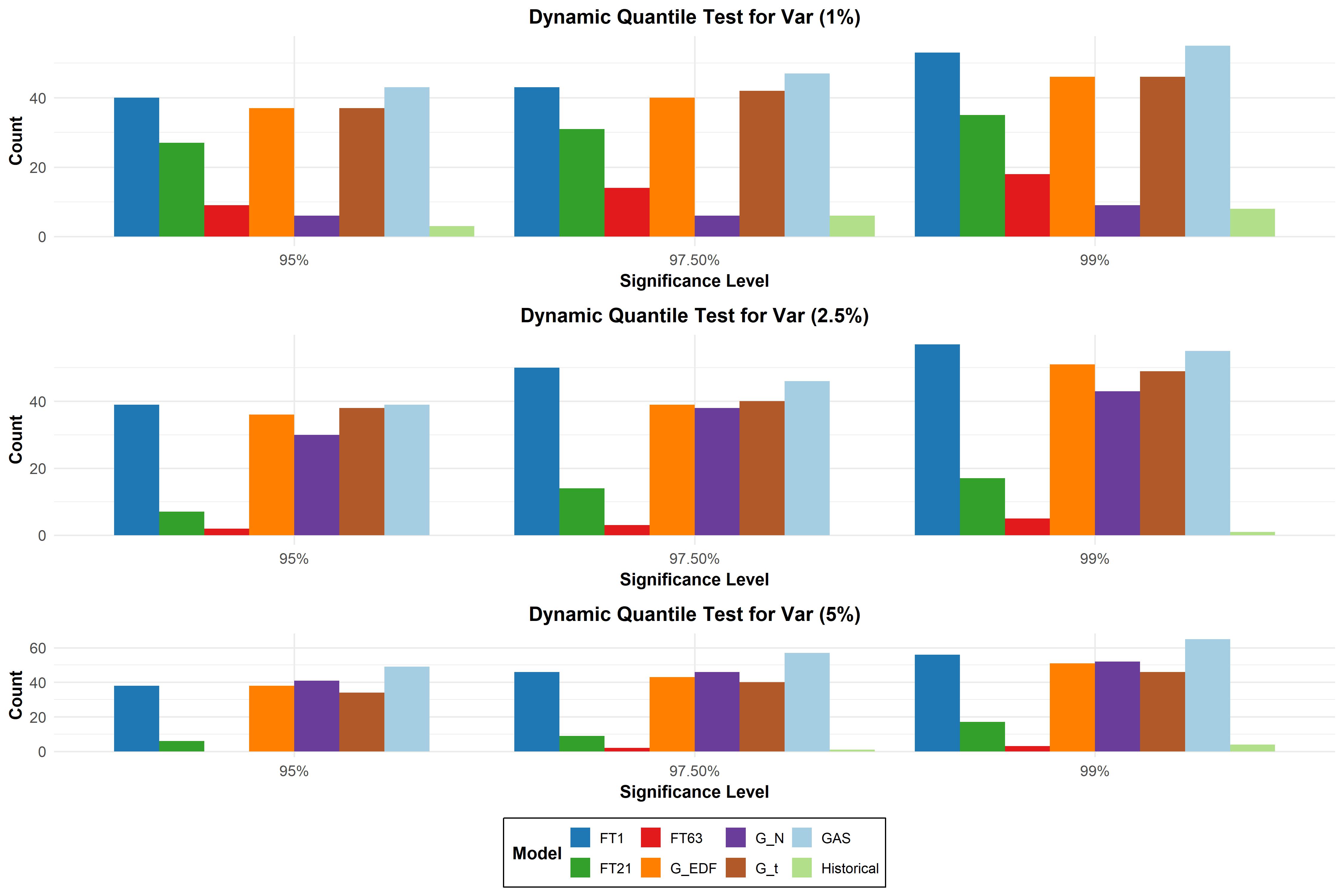}
\quad
\includegraphics[scale=0.30]{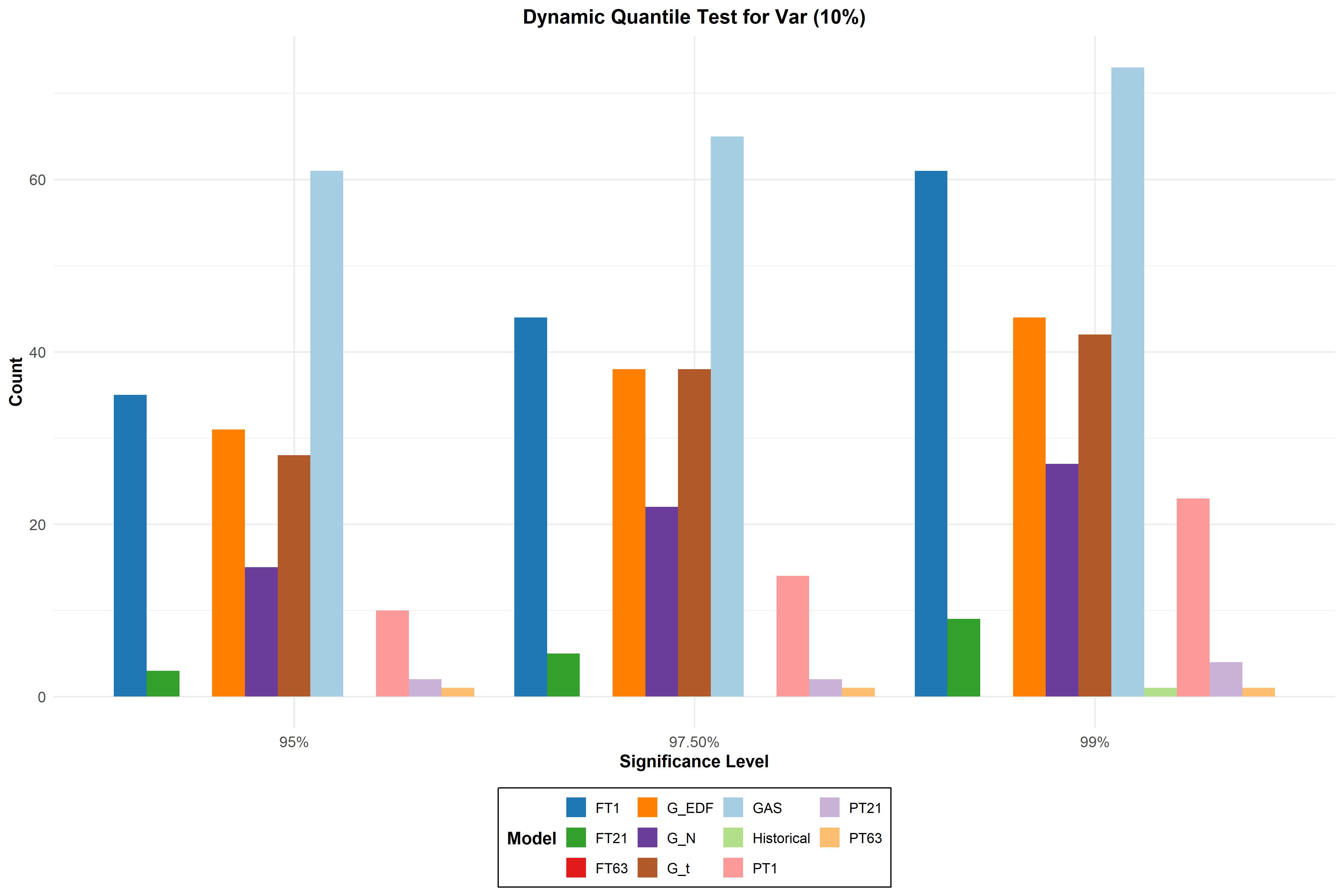}
\caption{The bar plot of the number of assets (out of 92) for which we failed to reject the null hypothesis in the Dynamic Quantile (DQ) test, at various value-at-risk (VaR) levels, on out-of-sample forecasts for the period January 2015 to September 2023. A higher number indicates better model performance, as it reflects a closer alignment between observed and expected VaR violations for each specified model. Results are reported for four VaR confidence levels (1\%, 2.5\%, 5\%, and 10\%) across different models.}
\label{dqplots}
\end{center}
\end{figure}

\end{document}